\documentstyle[12pt,bezier]{article}
\newcommand{\nc}{\newcommand}
\nc{\beq}{\begin{equation}}
\nc{\eeq}{\end{equation}}
\nc{\sss}{\scriptscriptstyle}
\def\rd{{\rm d}}
\begin{document}

\begin{titlepage}
\pagestyle{empty}
\baselineskip=21pt
\rightline{UMN-TH-1213-93}
\rightline{TPI-MINN-93/48-T}
\rightline{hep-ph/9401208}
\rightline{December 1993}
\vskip .2in
\begin{center}
{\large{\bf Protecting the Primordial Baryon Asymmetry \\
{}From Erasure by Sphalerons}} \end{center}
\vskip .1in
\begin{center}
James M.~Cline

Kimmo Kainulainen

and

Keith A. Olive

{\it School of Physics and Astronomy, University of Minnesota}

{\it Minneapolis, MN 55455, USA}

\vskip .2in

\end{center}
\vskip 1in
\centerline{ {\bf Abstract} }
\baselineskip=18pt
If the baryon asymmetry of the universe was created at the GUT scale,
sphalerons together with exotic sources of $(B-L)$-violation could have erased
it, unless the latter satisfy stringent bounds.  We elaborate on how the small
Yukawa coupling of the electron drastically weakens previous estimates of these
bounds.
\noindent
\end{titlepage}
%\newpage
\baselineskip=18pt
{\newcommand{\la}{\mbox{\raisebox{-.6ex}{$\stackrel{<}{\sim}$}}}
{\newcommand{\ga}{\mbox{\raisebox{-.6ex}{$\stackrel{>}{\sim}$}}}

\section{Introduction}

The first proposals to explain the baryon asymmetry of the universe suggested
that baryogenesis occurred at the very high temperatures characteristic of the
scale of grand unification.  While there has been great interest in generating
baryons at the electroweak scale, the feasibility of this approach remains
uncertain, whereas baryon-production in grand unified theories can be
calculated fairly reliably in the framework of a given GUT. However there are
many potential opportunities for a baryon asymmetry produced at $10^{16}$ GeV
to be destroyed later on, which leads to interesting constraints on the
coefficients of operators that violate baryon or lepton number.

A baryon asymmetry generated at the GUT scale will survive only if baryon
number violating interactions are subsequently out of thermal equilibrium. In
most $SU(5)$ grand unification schemes, any baryon asymmetry generated will be
accompanied by a lepton asymmetry such that $B - L = 0$. Non-perturbative
electroweak interactions via sphalerons also violate baryon and lepton
(actually $B + L$) number at high temperatures \cite{krs} while perserving $B -
L$.  As such, the GUT-scale produced baryon asymmetry will be erased \cite{am}
if sphaleron interactions are in thermal equilibrium, as they are expected to
be at temperatures between $\sim 10^2$ and $10^{12}$ GeV.  By the same token,
any GUT-scale baryon asymmetry with a net $B - L$ asymmetry will be preserved.

If in addition to the sphaleron interactions there was another baryon or lepton
number violating interaction in a linear combination other than $B + L$ and in
thermal equilibrium, all net baryon and lepton numbers would be erased. Such
was the case encountered by Fukugita and Yanagida (FY) \cite{fy2} when
considering a simple extension to the standard electroweak model to allow for
Majorana neutrino masses. In order to maintain a net baryon asymmetry, this
source of lepton number violation is constrained to be out of equilibrium at
the times when sphalerons are in thermal equilibrium. This constraint was then
translated into a bound on the neutrino mass.

FY derived the conservative bound $m_\nu \ga 50$ keV, assuming only that
sphalerons are in equilibrium at $T \sim 10^2$ GeV. It was subsequently argued
that this constratint could be made substantially stronger because sphalerons
are in equilibrium at significantly higher temperatures as well \cite{ht}.
These arguments were also extended to other sources of baryon and lepton number
violation \cite{cdeo12,sonia}.

In a previous paper \cite{cko2} we pointed out that the constraints are
dramatically weakened when one realizes that right-handed electrons can protect
the baryon asymmetry in cases where it would otherwise have been destroyed, so
long as the interactions of the operators in question fall out of equilibrium
before the interactions of $e_R$ with $e_L$ and Higgs bosons come {\it into}
equilibrium. The idea that the weakness of $e_R$-$e_L$-Higgs interactions might
have something to do with preserving the baryon asymmetry from sphalerons and
other $B$- or $L$-violating effects was originally considered by the authors of
ref.~\cite{cdeo3} (CDEO), for the initial condition that $B-L=0$.  Their goal
was to determine whether $e_R \leftrightarrow e_L$ equilibrium might be
established so late that sphaleron interactions would not have enough time to
nullify the baryon asymmetry.

The argument for the importance of $e_R$ decoupling  is intuitively very simple
and was presented in a rudimentary form in our previous work \cite{cko2}.
Since we feel that the idea has far-reaching consequences, it seems worthwhile
to rederive the earlier results in a more accurate and rigorous fashion.
Moreover there is a variety of special cases for the initial conditions or the
depletion at intermediate scales of the $B$ and $L$ asymmetries, each with its
own subtleties, the discussion of which was beyond the scope of the first
paper.   Filling these gaps is the goal of the present work.

In what follows we will develop the full network of Boltzmann equations
necessary to solve for the baryon asymmetry through the thermalization of
$e_R$. In section 3, we derive the rates for decays and inverse decays as well
as scatterings to determine the $e_R$ decoupling temperature. We then consider
the case with $B - L = 0$ as did CDEO, in section 4. Although their result
indicated that the baryon asymmetry was not preserved in this case, the final
baryon asymmetry is exponentially sensitive to changes in reaction rates, and
we have undertaken a careful reanalysis of some effects which were treated too
roughly by CDEO.  In particular, we have included the effects of scatterings.
Qualitatively, our results confirm those earier conclusions. In section 5, we
generalize to the case where $B - L \ne 0$ and describe several mechanisms in
which a baryon asymmetry can be preserved from sphaleron effects.  Our
conclusions will be given in section 6.

\def\er{{e_{\sss R}}}
\def\el{{e_{\sss L}}}
\def\rd{{\rm d}}
\def\lra{\leftrightarrow}
\section{The network of Boltzmann equations}

In this paper we explore the consequences of the fact that the right-handed
chiral electrons $e_R$ remain out of equilibrium to suprisingly low
temperatures in the early universe. This phenomenon arises simply due to the
smallness of the electron mass or its Yukawa coupling, $h_e \simeq 2.94 \times
10^{-6}$. At temperatures above the electroweak phase transition electrons are
massless, or more accurately, different chiral states have only
chirality-conserving temperature-dependent masses, so that the only processes
that change lepton chirality are interactions with Higgs bosons.\footnote{ For
the quarks, QCD instanton interactions also violate chirality, but it is easy
to show that these give no new conditions on the chemical potentials; they are
consistent with the constraints associated with Higgs boson interactions.} The
lepton chirality-changing rate is therefore proportional to the small number
$h_e^2$. The smallness of the $e_R$-$e_L$ equilibration temperature $T_*$ is
important because the equilibrium conditions in the primieval plasma are
different above and below $T_*$. This has far-reaching consequences for the
regeneration of the baryon asymmetry and leads to a considerable relaxation of
bounds imposed on various exotic interactions to preserve the primordial baryon
asymmetry.

It was noted in CDEO that in the particular case of a universe with initial
$B-L=0$ the final baryon asymmetry is exponentially sensitive to the value of
the rate of the chirality changing interactions. In such a case special care
must be taken in computing the thermally averaged rates and also in deriving
the evolution equations for the asymmetries from the Boltzmann equations for
the particle densities. Indeed the evolution equations found in the literature
are often somewhat inaccurate in their treatment of the relevant collision
terms in the Boltzmann equations.

In this section we will carefully derive evolution equations for the particle
asymmetries in the early universe during the time when $e_R \lra e_L$
transitions come into equilibrium. In general these equations couple all
particle species present in the heat bath and involve a very large number of
reaction rates. Fortunately a natural hierarchy exists between different
reactions; various interaction processes can be divided into `fast' or `slow'
depending on how their rates compare to the Hubble expansion rate of the
universe $H$ at the timescale of interest. A process is called fast only if its
rate is much bigger than $H$, otherwise it is considered to be slow. Fast
processes need not be treated in detail, but instead, their effect can be
accounted for by an appropriate set of equilibrium conditions between chemical
potentials and certain conservation laws that restrict the explicit equations
governing the slow processes.

Examples of fast processes are the various electroweak interactions connecting
fermions in the weak isospin doublets, $W^- \lra \bar{\nu_e}e^-, \bar{u}d$,
etc., and the interactions  mediated by the electroweak anomaly, i.e.\ the
sphaleron interactions \cite{krs}. Naturally, all interactions that change the
chirality of electrons are taken to be slow, but there can also be others, due
to new physics. In Table 1 we have collected all slow processes that are
relevant to our considerations in this section. For illustrative purposies we
also show an electron lepton number violating interaction, which we will study
in detail in section 5 below.

\begin{table}
\begin{center}
\begin{tabular}{|l|c|} \hline\hline
(Inverse) Decays
& $H^* \leftrightarrow e_R{L_e}^c$ \\ \hline
Gauge scattering
& $e_RH\leftrightarrow L_eA$  \\ \cline{2-2}
& $e_RA \leftrightarrow L_eH^*$  \\ \cline{2-2}
& $e_R{L_e}^c \leftrightarrow AH^*$  \\ \hline
Fermion scattering
& $e_RL_f \leftrightarrow L_ef_R$  \\ \cline{2-2}
& $e_R{L_f}^c \leftrightarrow {L_e}^cf_R$ \\ \cline{2-2}
& $e_R{L_e}^c \leftrightarrow {L_f}^cf_R$  \\ \hline\hline
Electron lepton & \\
number violation
& $e_LL_\ell \leftrightarrow h^-H$  \\ \hline\hline
\end{tabular}
\end{center}
\noindent Table 1: Shown are all tree level interactions that change the
chirality of electrons and also a particular type of electron lepton number
violation. $H$ is the Higgs doublet, with components $(h^+,h)^{\sss T}$. $L_i$
refers to either of the members in a left-handed doublet; $f$ represents all
fermions but $\ell$ is restricted to the leptons. $A$ can be either the $B$ or
the $W$ gauge field.
\end{table}

Evolution equations for the asymmetries in the $e_R$ and $e_L$ species can be
written down in a quite general manner, but to make the equations close
typically requires some additional constraint that is specific to a given
system. The problem shows up, for example, in the appearance of the Higgs
chemical potential $\mu_0$ in the evolution equations; the expression for
$\mu_0$ in terms of $\mu_{e_R}$ and  $\mu_{e_L}$, if it can be given, depends
on the particular model.  It is also clear that if any particle number other
than $e_L$ or $e_R$ numbers was softly violated by some new exotic {\it slow}
interaction, then additional equations governing the evolution of the
corresponding asymmetry would be needed. Here we will restrict ourselves to the
case where at most $e_L$ violation (in addition to $e_R$-$e_L$ equilibration)
is tracked in detail, which includes the special situations that we study in
sections 4 and 5. Our derivation however will address all the subtleties needed
to derive evolution equations for any other system of interest.

\subsection{Equilibrium conditions}

Let us first consider some of the equilibrium conditions established by the
fast processes. Because sphaleron interactions fall out of equilibrium very
quickly below the electoweak phase transition temperature \cite{Shapo} and
because the sphaleron reprocessing of the changing $e_R$ asymmetry is crucial
for our considerations in this paper, it is sufficient for us to consider the
evolution equations only at temperatures above the phase transition. In the
high temperature phase, the electroweak interactions mediated by the charged
gauge bosons and scalars force the chemical potentials of quarks and leptons to
obey the equilibrium conditions \cite{ht}
\begin{eqnarray}
\phantom{\mu_{u_R}-}\mu_{d_L} = \mu_{u_L} \hspace{3pt}
&\phantom{maja}&
\phantom{\mu_{\ell_R}m}\mu_{e_L} = \mu_{\nu_{e}} \nonumber \\
\hspace{-1pt}\mu_{u_R} - \mu_{u_L} = \mu_0 \phantom{ -} &\phantom{maja}&
\phantom{\mu_{\ell_R}m}\mu_{\ell_L} = \mu_{\nu_{\ell}}\phantom{m}
\label{chem} \\
\hspace{5pt}\mu_{d_R} - \mu_{d_L} =-\mu_0&\phantom{maja}&
\hspace{-4pt}\mu_{\ell_R}-\mu_{\ell_L} = -\mu_0.
\phantom{-}
\nonumber
\end{eqnarray}
Here the labels $u$ and $d$ refer to all quarks with weak isospin $+1/2$ and
$-1/2$ respectively, and $\ell$ refers to the muon and tau lepton; the
chirality equilibrium equation is not assumed to hold for electrons. $\mu_0$ is
the chemical potential of the Higgs field. We have assumed that the universe
was initially isospin-neutral,\footnote{In contrast to the charge neutrality of
the Universe, which is proved to a very high degree by observations, there is
no requirement for the universe initially to have had zero net isospin $Q_3$;
any primordial isospin would have vanished immediately after the phase
transition leaving no observable trace today.  However, it is improbable that
such an asymmetry could have been dynamically generated because $Q_3$
corresponds to a gauge symmetry. Moreover, any initial value for it would have
been washed away by inflation. We therefore adopt the value $Q_3 = 0$ as the
most natural choice, albeit not an absolutely necessary.} implying that the $W$
chemical potential, which would otherwise appear in eq.~(\ref{chem}) above, is
zero at high temperatures \cite{ht}. Note also that due to the flavour mixing
in the quark sector all quark flavours have the same chemical potential.
Equilibrium sphaleron interactions impose a further condition between the
chemical potentials of quarks and leptons
\begin{equation}
9 \mu_{u_L} + \mu_{e_L} + \sum_{\ell=\mu,\tau} \mu_{\ell_L} = 0,
\label{SphEQ}
\end{equation}
and the charge neutrality of the universe leads to
\begin{equation}
Q = 6\mu_{u_L} + 13\mu_0 - \mu_{e_L} - \mu_{e_R}
        - 2\sum_{\ell=\mu,\tau} \mu_{\ell_L}= 0.
\label{charge}
\end{equation}
Equations (\ref{SphEQ}--\ref{charge}) contain six unknown parameters. In
addition to (\ref{SphEQ}--\ref{charge}) we will always have four additional
conditions on the chemical potentials so that the system has a unique solution.
Assuming that there were no new interactions, then well above $T_*$ we would
have the following conservation laws:\footnote{We warn the reader that the
symbol $L_e$ can mean the electron lepton doublet field, the asymmetry in
electron lepton number, or the U(1) symmetry for electron lepton number,
depending on the context. A similar warning applies for $L_\mu$ and $L_\tau$.}
\begin{eqnarray}
\frac{1}{3}B - L_e &=& C_{e,p} \nonumber \\
\frac{1}{3}B - L_\mu &=& C_{\mu,p} \nonumber \\
\frac{1}{3}B - L_\tau &=& C_{\tau,p} \nonumber \\
L_{e_R} &=& C_{e_R,p},
\label{claws}
\end{eqnarray}
where $L_f \equiv L_{f_L}+L_{f_R}+L_{\nu_{f}}$ and the subscripts $p$ in the
constants $C$ indicate that they are the primordial values. The connection
between the asymmetries and the chemical potentials is
\begin{equation}  L_f \equiv
        \frac{n_f-n_{f^c}}{s} \simeq  \frac{15c}{4\pi^2g_*} \frac{\mu_f}{T}
        \ +\ O((\mu_f/T)^3),
\label{asymmetry}
\end{equation}
where $c$ is 1 for fermions and 2 for bosons and $s=\frac{2\pi^2}{45}g_*(T)T^3$
is the entropy density of the universe.

If any of the conservation laws (\ref{claws}) are violated by exotic fast
interactions, then they must be replaced by new equilibrium conditions. Indeed,
the situation well below $T_*$ provides an example of how this works.  There
the $e_L-e_R$ transitions violate $e_R$ number strongly, so that the last
equation in (\ref{claws}) is replaced by the equilibrium condition $\mu_{e_R} -
\mu_{e_L} = - \mu_0$. On the other hand, if some of the conservation laws are
softly violated, so that around $T_*$ the rates of relevant interactions are
comparable to the expansion rate, then explicit equations must be written for
the asymmetries affected by these interactions. In this section we will
concentrate on the cases where $L_{e_R}$ is violated by itself or in
combination with soft electron lepton number violation.

\subsection{The evolution equations}

To derive the evolution equations for the $e_R$ and $e_L$ asymmetries we first
write down the Boltzmann equations for the difference between the particle and
antiparticle number densities. The collision terms in these equations
correspond to the interactions shown in the table 1 and depicted in figure 1.
It is reasonable to assume that each particle is in kinetic equilibrium at all
times of interest; this is true in particular for the right-handed electrons
due to their nonvanishing hypercharge. We then find
\begin{eqnarray}
(\partial_t - 3H)(n_{e_R}-n_{{e_R}^c})&=&
\hspace{0.5em} 2{\displaystyle \int}
{\displaystyle \prod_{i=1}^3} \rd \Pi_i (2\pi )^4\delta^4(p_1-p_2-p_3)
|{\cal M}(h^*\leftrightarrow {e_L}^ce_R)|^2   \nonumber \\
&&\hspace{2.5em} \times \left\{  f_{h^*}(1-f_{{e_L}^c})(1-f_{e_R})
                  - f_{{e_L}^c}f_{e_R}(1+f_{h^*}) \right\} -  {\rm c.c.}\
\nonumber \\
&&\hspace{-0.3em}+2{\displaystyle \int}{\displaystyle \prod_{i=1}^4} \rd
\Pi_i \, (2\pi)^4\delta^4(p_1+p_2-p_2-p_3)  \sum_f \left\{|{\cal
M}({f_L}^cf_R \leftrightarrow {e_L}^ce_R)|^2  \right. \nonumber \\
& &\hspace{2.2em} \times [f_{{f_L}^c}f_{f_R}(1-f_{{e_L}^c})(1-f_{e_R})
           - f_{{e_L}^c}f_{e_R}(1-f_{{f_L}^c})(1-f_{f_R}) ]
      \nonumber \\
& &\hspace{2.7em} \left.  + \cdots \mbox{} \right\} - {\rm c.c.}\
\label{BE01a}
\end{eqnarray}
\begin{eqnarray}
(\partial_t - 3H)(n_{e_L}-n_{{e_L}^c})  &=& \hspace{-0.5em}
 - \frac{1}{2}(\partial_t - 3H)(n_{e_R}-n_{{e_R}^c}) \nonumber \\
& &\hspace{-0.5em}+{\displaystyle \int} \displaystyle{\prod_{i=1}^4}
\rd \Pi_i \, (2\pi)^4\delta^4(p_1+p_2-p_2-p_3)
 \sum_{\ell=e,\mu,\tau}
 \left\{ S_\ell |{\cal M}( h^-h^-\leftrightarrow \ell_Le_L)|^2 \right.
\nonumber \\
& &\hspace{1.7em}\times [ f_{h^-}f_{h^-}(1-f_{e_L})(1-f_{\ell_L})
        - f_{e_L}f_{\ell_L}(1+f_{h^-})(1+f_{h^-}) ] \nonumber \\
& &\hspace{1.7em}\left. + \cdots \right\} -  {\rm c.c.}
	+ \frac{1}{6}(\partial_t - 3H)(n_B - n_{B^c})
%{\rm F.P.}\
\label{BE01b}
\end{eqnarray}
where $H$ is the Hubble expansion rate and we defined the usual phase space
volume element as $\rd \Pi \equiv \rd^3p/(2\pi)^32E$. $f_L$ and $f_R$ are the
chiral quarks or leptons. The symmetry factor $S_\ell$ in (6) is introduced to
avoid double-counting in case of identical initial or final state particles:
$S_e=\frac{1}{4}$ and $S_{\mu,\tau}=\frac{1}{2}$. According to our assumption
of prevailing kinetic equilibrium, the distribution functions are the usual
Fermi-Dirac or Bose-Einstein functions
\begin{equation}
f_i(p,\mu_i) = (e^{\beta(E_i-\mu_i)}\pm 1)^{-1},
\label{distribution}
\end{equation}
where $\beta \equiv 1/T$ and $T$ is the common temperature. The factors of 2 in
front of the integrals in (\ref{BE01a}) come from summing over the equal
contributions from both fields in each left handed doublet.  This is evident
because the couplings and therefore the temperature-dependent masses of the
doublet components are identical. Similarly, the half in front of the first
term in (\ref{BE01b}) arises because $e_L$-number is changed only in half of
these interactions.

In eqs.~(\ref{BE01a}--\ref{BE01b}) we explicitly wrote only a few
representatives of the whole set of collision terms corresponding to the
interactions listed in the table 1, the rest being denoted by ellipses. The
${\rm c.c.}$ terms that come from the evolution equation for $n_{f^c}$ are just
the charge conjugates of the terms shown, and the term
\beq
\frac{1}{6}(\partial_t - 3H)(n_B-n_{B^c}) \equiv \frac{1}{6}s\dot B
\label{Bdot}
\eeq
in eq.~(\ref{BE01b}) is due to the fast interactions of sphalerons, which
change $\el$ number at $1/6$ the rate at which they violate baryon number. (In
eq.~(\ref{Bdot}) we used the fact that $\dot{s}/s=-3H$ in the adiabatically
expanding universe.)  To see that this term is necessary, recall that in the
absence of the $\Delta L=2$ lepton-violating interaction, we have the conserved
quantity $\frac{1}{3}B-L_e = \frac{1}{3}B - 2L_\el - L_\er$. If the $\Delta
L=2$ scattering term in (\ref{BE01b}) is zero, we see that this conservation
law will be satisfied only if the $\frac{1}{6}s\dot B$ term is present.  Below
we will express $B$ in terms of $L_\er$ and $L_\el$. Lastly, let us remark that
the second term in (\ref{BE01b}) is particular to the model with electron
lepton number violation and should be omitted when $\frac{1}{3}B-L_e$ is
conserved.

Equations (\ref{BE01a}--\ref{BE01b}) constitute a set of nonlinear equations
for the chemical potentials that appear in the distribution  functions
$f_i(p,\mu_i)$. We can convert them into linear equations for the asymmetries
by expanding the collision terms to first order in the $\beta \mu_i$, which are
proportional to the asymmetries through eq.~(\ref{asymmetry}). Of course the
resulting evolution equations are only valid for small chemical potentials, but
this is a reasonable sacrifice because we are interested only in very small
final asymmetries of the order of the present baryon asymmetry $\sim 10^{-10}$.
Even in the case of a large initial asymmetry the nonlinearity would persist
only for a very short time, after which the system would be well-described by
the linear equations.  Then performing the expansions in chemical potentials
and using equations (\ref{chem}) to rewrite the chiral fermion asymmetries in
terms of $\mu_0$, equations (\ref{BE01a}-{\ref{BE01b}) become
\begin{eqnarray}
\frac {\rd L_{e_R}}{\rd t} &=& - \frac{2\beta}{s}
(\mu_{e_R}-\mu_{e_L}+\mu_0) \times (I_D + I_G + \sum_f I_f )
\nonumber \\
\frac {\rd L_{e_L}}{\rd t} &=& - \frac{1}{2}\frac{\rd L_{e_R}}{\rd t}
- \frac{2\beta}{s} \sum_{\ell=e,\mu,\tau}
(\mu_{e_L}+\mu_{\ell_L}+2\mu_0) I_{\Delta L,\ell} + \frac{1}{6}\dot B
\label{BEM}
\end{eqnarray}
It should be noted that the chirality-changing part of the collision term in
eq.~(\ref{BEM}) for $L_{e_R}$ is proportional to $\mu_{e_R}-\mu_{e_L}+\mu_0$,
clearly showing that $L_{e_R}$ is driven towards the correct equilibrium value.
These equations are not yet solvable because $\mu_0$ and the sphaleron term
$\frac{1}{6}\dot B$ appearing on the r.h.s.~of (\ref{BEM}) must be written in
terms of the asymmetries $L_{e_R}$ and $L_{e_L}$. We will return to this issue
shortly.

The coefficients that multiply chemical potentials in (\ref{BEM}) are defined
as follows.  The (inverse) decay process gives
\begin{equation} I_D =  2
{\displaystyle \int} \, {\displaystyle \prod_{i=1}^3} \rd \Pi_i \, (2\pi
)^4\delta^4(p_1-p_2-p_3)  |{\cal M}({e_L}^c e_R  \leftrightarrow h^*)|^2
f_{e1}^0f_{e2}^0(1+f_{h3}^0).
\label{ID}
\end{equation}
Scatterings involving the gauge bosons\footnote{We thank Sonia Paban and Willy
Fischler for originally pointing out this contribution to us.} contribute a
term
\begin{eqnarray}
I_G = 2 {\displaystyle \int }\!\!
&{\displaystyle \prod_{i=1}^4}& \!\! \rd \Pi_i \, (2\pi
)^4\delta^4(p_1+p_2-p_3-p_4) \times \nonumber \\ &\times & \left\{
|{\cal M}({e_L}^ce_R \leftrightarrow h^*A)|^2 f_{e1}^0 f_{e2}^0
(1+f_{h3}^0) (1+f_{A4}^0)
\right. \nonumber \\  &+& \: \left.
\hspace{3pt}|{\cal M}(Ae_R \leftrightarrow h^*e_L)|^2
f_{e1}^0f_{A2}^0(1-f_{e3}^0)(1+f_{h4}^0) \right. \nonumber \\
&+& \: \left.
\hspace{3pt}|{\cal M}(he_R \leftrightarrow Ae_L)|^2
f_{e1}^0f_{h2}^0(1-f_{e3}^0)(1+f_{A4}^0)  \right\},
\label{Ih}
\end{eqnarray}
where $A$ refers to either of the vector bosons $W$ or $B$. The scatterings
involving fermions give
\begin{eqnarray}
I_f = 2 {\displaystyle \int} \!\!
&{\displaystyle \prod_{i=1}^4}& \!\! \rd \Pi_i \, (2\pi
)^4\delta^4(p_1+p_2-p_3-p_4) \times \nonumber \\ &\times& \left\{
|{\cal M}({e_L}^ce_R)\leftrightarrow {f_L}^cf_R) |^2
f_{e1}^0 f_{e2}^0(1-f_{f 3}^0) (1-f_{f 4}^0) +
\right. \nonumber \\
&+& \mbox{} (\: |{\cal M}(f_L e_R \leftrightarrow f_R e_L)|^2
+ |{\cal M}({f_R}^c e_R \leftrightarrow {f_L}^c e_L)|^2 \:) \times
\nonumber \\
& & \left. \mbox{} \times
 [f_{e1}^0f_{f 2}^0(1-f_{e3}^0)(1-f_{f 4}^0)] \right\}.
\label{If}
\end{eqnarray}
Finally, the lepton number violating collision term is
\begin{eqnarray}
I_{\Delta L,\ell} = {\displaystyle \int} \!\!
&{\displaystyle \prod_{i=1}^4}& \!\! \rd \Pi_i \, (2\pi
)^4\delta^4(p_1+p_2-p_3-p_4) \times \nonumber \\
&\times&
\left\{ (S_\ell |{\cal M}(e_L\ell_L\leftrightarrow h^-h^-)|^2
 + |{\cal M}(e_L \nu_{\ell}\leftrightarrow h^*h^-)|^2) \times \right.
\nonumber \\
& & \hspace{1em} \left. \times f_{e1}^0 f_{\ell 2}^0
(1+f_{h3}^0) (1+f_{h 4}^0) \right.  \nonumber \\
& &
\left. + \; (|{\cal M}(e_Lh^+\rightarrow {\ell_L}^ch^-)|^2
 + |{\cal M}(e_L h^+\rightarrow {\nu_{\ell}}^ch^*)|^2
 + |{\cal M}(e_L h\rightarrow {\nu_{\ell}}^ch^-)|^2) \times \right.
\nonumber \\
& & \left. \hspace{1em} \times f_{e1}^0 f_{h 2}^0
(1-f_{\ell 3}^0) (1+f_{h 4}^0) \right\}.
\label{IL}
\end{eqnarray}
The distribution functions $f^0_{\alpha i}$ appearing in (\ref{ID}--\ref{IL})
are the usual equilibrium functions defined in (\ref{distribution}) for
particle species $\alpha$ with zero chemical potential and momentum $p_i$.

There is a technical subtlety concerning the $s$-channel scattering
contributions to the chirality-changing interactions.  We have included the
(inverse) decay contribution explicitly in our Boltzmann equations, which
assumes that the Higgs particle is a thermalized member of the heat bath.
However, the on-shell part of the Higgs exchange in the $s$-channel scattering
physically corresponds to an inverse decay into a real Higgs boson followed
immediately by   a decay process. Thus the on-shell contribution in $s$-channel
scattering introduces double counting of decay interactions and  has to be
subtracted away. Hence what we mean by the $s$-channel contributions in our
Boltzmann equations is actually the on-shell subtracted part. We will discuss
the subtraction procedure in section 3, along with the computations of matrix
elements and thermal averages.

We still have the problem that in general $\mu_0$ is not simply related to
$\mu_{e_R}$ and $\mu_{e_L}$. This would be the case for example if all the
lepton numbers were violated by slow processes. Then four conditions would be
needed to close the network of equations for the four softly-violated
asymmetries \cite{cko1}. But if we assume that only electron lepton number
violation is slow, we can use the conservation laws for $\frac{1}{3}B-L_\mu$
and $\frac{1}{3}B-L_\tau$ (or the new chemical equilibrium conditions for
$\mu_{\mu_L}$ and $\mu_{\tau_L}$ if these are violated) to express the
remaining unknown chemical potentials in (\ref{BEM}) in terms of $\mu_{e_R}$
and $\mu_{e_L}$.  In particular we can parametrize $\mu_0$ as
\begin{equation}
        \mu_0 \equiv a\mu_{e_R} + b\mu_{e_L}
\label{muHiggs}
\end{equation}
with $a$ and $b$ model-dependent constants, computable when the specific set of
conservation laws and equilibrium conditions is given.

Using (\ref{SphEQ}--\ref{charge}) and (\ref{muHiggs}) one then finds
\begin{equation}
        B = \frac{1}{2}(1-13a)L_{e_R} - \frac{1}{2}(1+13b)L_{e_L}.
\label{Bequ}
\end{equation}
By differentiating this with respect to time, we can
write $\frac{1}{6}\dot B$ in terms of $\dot L_\er$ and $\dot L_\el$ in
eqs.~(\ref{BEM}), so that they close to give
\begin{eqnarray}
\frac {\rd
L_{e_R}}{\rd t} = &-& \Gamma_{RL} ((1+a)L_{e_R}-(1-b)L_{e_L}) \nonumber \\
\frac {\rd L_{e_L}}{\rd t}= &-& \frac{5+13a}{13(1+b)}\frac{\rd L_{e_R}}{\rd t}
- \sum_{\ell=e,\mu,\tau}\Gamma_{\Delta L,\ell }
(a_\ell L_{e_R}+b_\ell L_{e_L}+c_\ell),
\label{BEF}
\end{eqnarray}
where
\begin{eqnarray}
\Gamma_{RL} &\equiv& \frac{12}{T^3}(I_D + I_G + \sum_f I_f);
\nonumber \\
\Gamma_{\Delta L,\ell} &\equiv& \frac{12}{T^3} I_{\Delta L,\ell}.
\label{Gamma}
\end{eqnarray}
In (\ref{BEF}), the coefficients $a_\ell$ and $b_\ell$ are linear combinations
of $a$ and $b$, and $c_\ell$ are related to the initial values of conserved
quantities.  For example, if there is fast violation of muon and tau lepton
number, then $\mu_{\mu_L} = \mu_{\tau_L} = -\mu_0$, as will be shown in section
5, and we obtain
\begin{eqnarray}
        & a_e = \frac{24a}{13(1+b)}
        & \qquad b_e = \frac{24}{13} \nonumber \\
        & a_\mu = a_\tau =\frac{1}{2}a_e
        & \qquad b_\mu = b_\tau = \frac{1}{2}b_e\nonumber \\
        & c_i = 0. &
\label{example}
\end{eqnarray}
Nevertheless, in equations (\ref{BEF}--\ref{Gamma}) the parameters $a$, $a_i$,
$b$, and $b_i$ are left unspecified in order to stress the fact of their
model-dependence, notably their dependence on whatever boundary conditions are
imposed by the fast processes.   Since these appear multiplying the rates, the
temperature at which the asymmetries reach their asymptotic values can be
accurately determined only after these parameters are known.

If there is no lepton number violation aside from sphaleron effects, that is if
$\Gamma_{\Delta L} = 0$, the equation for $\dot{L}_{e_L}$ in (\ref{BEF}) is
actually redundant.  In that case one would use the conservation of
$\frac{1}{3}B - L_e$ along with eq.~(\ref{Bequ}) to express $L_{e_L}$ in terms
of $L_{e_R}$ in the evolution equation for $\dot{L}_{e_R}$, which could then be
solved for $L_{e_R}$ directly; this is exactly what we will do in section 4
below. If there does exist additional electron lepton number violation,
$\frac{1}{3}B - L_e$ is not conserved and the more general form of (\ref{BEF})
must be used.

\section{Rates for (inverse) decays and scatterings}

\def\er{{e_{\sss R}}}
\def\el{{e_{\sss L}}}
\def\elc{{{e_{\sss L}}^c}}
\def\tr{{t_{\sss R}}}
\def\tlc{{{t_{\sss L}}^c}}
\def\L{{\sss L}}
\def\H{{\sss H}}
\def\G{{\sss G}}
\def\ddot{\!\cdot\!}

We are interested in the rate at which $\er$ number violation occurs in the
early universe, to determine at what temperature the associated interactions
come into thermal equilibrium, and in particular whether this can be close to
the electroweak phase transition temperature. The dominant processes are
usually the decays and inverse decays of Higgs bosons into $\er$ and the lepton
doublet $L_e$ shown in figure 1.  As will be discussed in the next section, the
answer to the question of whether an $\er$ asymmetry relaxes completely before
sphalerons go out of equilibrium at the electroweak phase transition, is
exponentially sensitive to the total rate of $\er$ violation.  We have
therefore undertaken to more carefully estimate this rate than was done in
CDEO.  We will find that for very small Higgs boson masses $m_\H$, inverse
decays by themsleves are slow enough so that $\er$ violation would not take
place until after sphalerons had largely disappeared, in which case the baryon
asymmetry would survive despite the vanishing of $B-L$.  But before making this
conclusion one should consider corrections to the rate of $\er$ violation due
to two-body scattering processes.  It will be shown that these actually
dominate over the decays for small $m_\H$, leading to a large enough rate of
$\er$ violation that the results of CDEO are qualitatively unchanged.

\subsection{Decays}

For the (inverse) decay rate, we will make several refinements on the treatment
given by CDEO: inclusion of the thermal masses of the leptons, use of the
finite-temperature wave functions for the fermions, and retention of the
Bose-Einstein statistical factor for the Higgs boson.  The first two
corrections are important in the limit of small $m_\H$ because they exacerbate
the the phase space suppression for the decay.

The decay rate entering into the Boltzmann equation (\ref{BEM}) is proportional
to an integral $I_D$, eq.~(\ref{ID}), in which appears the transition matrix
element ${\cal M}$. In terms of the electron Yukawa coupling $h_e$, the squared
matrix element for $H\to\er\el^c$ is
\beq
\label{d2}
        |{\cal M}|^2 = 2h_e^2 {\hat p}_\el\ddot {\hat p}_\er =
        2h_e^2(|\vec p_\el||\vec p_\er|-\vec p_\el\ddot\vec p_\er),
\eeq
where ${\hat p}_\el$ and ${\hat p}_\er$ are light-like four-momenta constructed
from the corresponding three-momenta.  These appear in place of the usual
four-momenta because the fermion masses are from thermal effects which break
Lorentz invariance but preserve chirality, so that one must still use massless
spinors in constructing the amplitude \cite{weld}.  However to a good
approximation the dispersion relations for the particles appear to be massive.
The thermal masses of the two chiralities of electrons and the Higgs,
appropriate to the temperatures above the electroweak phase transition in which
we are interested, are given by\footnote{We note that the coefficient of
$m_0^2$ in the expression for $m_\H^2(T)$ is corrected here relative to that
which was quoted in ref.~\cite{cdeo3}.} \cite{weld,meg}
\begin{eqnarray}
        m_\el^2(T)/T^2 &=& (3g^2+g'^2)/32 \cong 0.044;\nonumber \\
        m_\er^2(T)/T^2 &=& g'^2/8 \cong 0.017;\nonumber \\
        m_\H^2(T)/T^2 &=& 2D(1-T^2_c/T^2);\nonumber \\
        D &\equiv & (2m^2_W+m^2_Z+2m^2_t+m^2_0)/8v^2.
\label{d3}
\end{eqnarray}
Here $m_0$ is the zero-temperature Higgs boson mass, and $g$ and $g'$ are the
respective SU(2)$_L$ and U(1)$_Y$ couplings evaluated at the running scale of
100 GeV which is at the lower limit of the temperatures we are interested in.
$T_c$ is the critical temperature for the electroweak transition, given at one
loop by
\beq
        T_c^2 = {1\over 4D}\left(m_\H^2 -{3\over 8\pi^2 v^2}(2m^4_W+m^4_z-
        4m^4_t) - {1\over 8\pi^2 v^4 D}(2m^3_W+m^3_Z)^2\right).
\label{d3a}
\eeq
By substituting the lowest experimentally allowable values of the top quark
mass $m_t=90$ GeV and $m_0 = 60$ GeV, we find that $m_\H/T$ must be no smaller
than $0.41$ in the high temperature limit, where it becomes constant as a
function of temperature. This also approximately coincides with the kinematic
threshold for thermal Higgs boson decays into electrons.  $m_\H/T$ is the key
parameter for the strength of the decays because $I_D$ depends entirely upon
it, considering $m_\el$ and $m_\er$ to be known quantities.  In Figure 2 we
have plotted the thermal Higgs mass for various values of $m_0$ and $m_t$,
where it can be seen that $m_\H$ varies between 0.4 and 2 for $m_0$ between 60
and 1000 GeV, for relevant values of the top quark mass.

The integral in eq.~(\ref{ID}), which is proportional to the thermally averaged
rate of decays, can be reduced to one dimension:
\beq
I_D = \frac{m_\H T^3h_e^2\gamma^2}{\pi^3}\int_1^\infty \rd u \frac{e^{um_\H/T}}
        {(e^{um_\H/T}-1)^2}
\! \ln \left( \frac {\textstyle \cosh (\alpha_\el u + \gamma \sqrt{u^2-1})}
               {\textstyle \cosh (\alpha_\el u - \gamma \sqrt{u^2-1})}
% \right. \nonumber \\ & \t \left.
\frac {\textstyle \cosh (\alpha_\er u + \gamma \sqrt{u^2-1})}
             {\textstyle \cosh (\alpha_\er u - \gamma \sqrt{u^2-1})} \right),
\label{d4}
\eeq
where
\begin{eqnarray}
\alpha_\el     &  \equiv & (m_\H^2+m_\el^2-m_\er^2)/4m_\H T; \nonumber \\
\alpha_\er     &  \equiv & (m_\H^2+m_\er^2-m_\el^2)/4m_\H T; \nonumber \\
\gamma         &  \equiv & \lambda^{1/2}(m_\H^2,m_\el^2,m_\er^2) /4m_\H T;
                                                                \nonumber\\
\lambda(x,y,z) &  \equiv & (x-y-z)^2-4yz.
\label{d5}
\end{eqnarray}
In CDEO, $I_D$ was approximated by $(h_e
m_\H T\ln 2)^2/(16\pi^3)$.  In Table 2 we compare the numerically integrated
value with this approximation for some values of $m_\H/T$.

\begin{table}
\def\t{\times}
\begin{center}
\begin{tabular}{|c|c|c|c|c|c|}
\hline \hline
$m_\H/T$   & $I_{D,CDEO}/K$ & $I_{D}/K$  & $I_{top}/K$ & $I_G/K$ \\ \hline
 0.4  &  1.6$\t 10^{-4}$  &  2.5$\t 10^{-5}$  &   $4.2\t 10^{-5}$  &  3.2$\t
10^{-4}$ \\
 0.6  &  3.5$\t 10^{-4}$  &  2.0$\t 10^{-4}$  &   $4.2\t 10^{-5}$  &  " \\
 0.8  &  6.2$\t 10^{-4}$  &  4.1$\t 10^{-4}$  &   $4.1\t 10^{-5}$  &  " \\
 1.0  &  1.0$\t 10^{-3}$  &  6.4$\t 10^{-4}$  &   $4.0\t 10^{-5}$  &  " \\
 1.2  &  1.4$\t 10^{-3}$  &  8.5$\t 10^{-4}$  &   $3.7\t 10^{-5}$  &  " \\
 1.4  &  1.9$\t 10^{-3}$  &  1.1$\t 10^{-3}$  &   $3.4\t 10^{-5}$  &  " \\
 1.6  &  2.5$\t 10^{-3}$  &  1.2$\t 10^{-3}$  &   $3.2\t 10^{-5}$  &  " \\
 1.8  &  3.1$\t 10^{-3}$  &  1.4$\t 10^{-3}$  &   $2.9\t 10^{-5}$  &  " \\
 2.0  &  3.9$\t 10^{-3}$  &  1.5$\t 10^{-3}$  &   $2.7\t 10^{-5}$  &  " \\
\hline\hline
\end{tabular}
\end{center}
\noindent Table 2:  The quantities $I_a$ that appear in the Boltzmann equation,
divided by $K = h_e^2 T^4$. The subscript CDEO refers to the approximations for
the decay rate made in reference \cite{cdeo3}, while $D$, $top$, and $G$ refer
respectively to the exact decay rate and the scattering rates for top quarks
and gauge bosons to produce $\er$.
\end{table}

\subsection{Scatterings producing $\er$}

One might expect that two-body scatterings which change $\er$ number are slower
than decays because of phase space and coupling constant suppression.  However
in the region of small thermal Higgs boson masses the phase space for the decay
is strongly suppressed.  In this regime the production of electrons by top
quark
scattering (Figure 3) or Higgs/gauge boson scattering (Figure 4) may actually
dominate over the decays.

In the Boltzmann equation, the thermally averaged scattering rates for
processes which change $\er$ number appear through the integrals $I_f$ and
$I_\G$ defined in eqs.~(\ref{Ih}--\ref{If}).  We may ignore all the fermion
scattering processes except for $f=top$ since the top quark has the dominant
Yukawa coupling. The channels which produce an $\er$ in the final state are
listed in Table 1, and are categorized here as top-quark or gauge/Higgs boson
scatterings.  We use the Maxwell-Boltzmann approximation for the initial state
particles, and neglect the distribution functions compared to unity for the
final state particles.  The errors introduced by this procedure are irrelevant
in the regime of large $m_\H/T$ where the scatterings are small corrections to
the decay rate; in the small $m_\H/T$ regime, although the scatterings
dominate, they are still sufficiently fast for our purposes that the
anticipated 20\% corrections to the Maxwell-Boltzmann approximation would not
change any of our subsequent conclusions.  Then the quantities $I_i$ can be
expressed as an integral over the Mandelstam variable $s$ involving the
corresponding cross sections and the modified Bessel function $K_1$:
\beq
\displaystyle I_i = {T\over 32\pi^4}\sum_c\int_{M_c^2}^\infty {ds\over s^{1/2}}
        \lambda(s,m_1^2,m_2^2) K_1(\sqrt{s}/T) \sigma_c(s)
\label{d6}
\eeq
The sum is over all possible channels $c$, each of which has a threshold
determined by the thermal masses of the initial or final state particles, $M_c
=
\max(m_1+m_2,m_3+m_4)$.  The sum includes a factor of 2 for isospin (since all
the processes we consider involve doublets) and if there are gauge bosons, a
factor of 3 for the polarizations.  Note that there is no corresponding factor
for the fermion spins because they all have a definite helicity.  The sum also
runs over the different crossings that leave $\er$ in the final state.  The
choice of whether $\er$ is in the initial or final state is arbitrary--either
way gives the same final answer.

In performing the thermal averages of the cross sections, we omitted
corrections of order $m^2/s$, where $m$ is a thermal mass, since typically
$m<T$ and $s\sim 20 T^2$.  Thus we can set the threshold to zero in the
integration of eq.~(\ref{d6}) and approximate $\lambda(s,m_1^2,m_2^2)$ by
$s^2$.  (The same neglect of thermal masses would not have been appropriate for
accurately determining the decay rate since the phase space there is determined
by the scale $m_\H$ rather than $s$.)  In this approximation, the analytic
expressions for the top quark scattering cross sections are
\beq
        \sum_c\sigma_c(s) = {(h_e h_t)^2\over 8\pi
         s} \left( {s^2\over (s-m_\H^2)^2 + s\Gamma^2} + 2  \right)
\label{d7a}
\eeq
and for gauge/Higgs boson scattering
\beq
        \sum_c\sigma_c(s) = {4h_e^2\over
         \pi sT^2} \left( m_\er^2\ln{s\over m_\er^2} + m_\el^2\ln{s\over
        m_\el^2} - {5\over 4}m^2_\el - {7\over 4}m^2_\er \right),
\label{d7b}
\eeq
where the thermal masses of eq.~(\ref{d3}) have been used and $\Gamma$ is the
width of a Higgs boson with mass $\sqrt{s}$:
        $\Gamma(s) \cong {h_t^2\over 16\pi}\sqrt{s}$.
Some care must be taken to obtain the subleading terms correctly.
The numerical values of the above contributions to
the Boltzmann equation for the $\er$ asymmetry are also shown in Table 2. We
have assumed a top quark Yukawa coupling corresponding to a mass of 90 GeV in
computing $I_{top}$.  Although this may be unrealistically small, we wanted to
examine the regime of small $m_\H/T$ which is most favorable to the possibility
of keeping $\er$-changing interactions out of equilibrium below the freeze-out
of sphalerons, and eq.~(\ref{d3}) shows that this occurs for the smallest
values of $m_t$ and $m_0$.  However, Table 2 shows that the contributions from
scattering dominate those of decays in the low-$m_\H$ region, so that one
cannot in fact evade the washout of an $\er$ asymmetry even by going to small
masses.  More will be said about this in section 4.

\subsection{Subtraction of Resonances}

As has been mentioned above, the $s$-channel scattering $\tr\tlc\to \er\elc$
contains an on-shell part which must be subtracted to avoid double-counting the
contribution from the inverse decays.  This necessity has been mentioned in
references \cite{kw} and \cite{fot}, although the details of how to carry it
out were not explained.  We have considered two methods that seem reasonable
and are in good quantitative agreement with each other. Writing the
finite-width, squared $s$-channel propagator for the Higgs as $P(s;\Gamma) =
((s-m^2_\H)^s + s\Gamma^2)^{-1}$, we can summarize the different methods by
making the following modifications to the propagator, which we refer to
respectively as the principal value (PV) and delta function subtraction (SDF)
methods:
\begin{eqnarray}
& {\rm PV:}\quad & P(s;\Gamma) \to P(s+i\epsilon;0); \nonumber \\
& {\rm SDF:}\quad & P(s;\Gamma) \to P(s;\Gamma) -
        {\displaystyle {\pi\over\sqrt{s}\Gamma}} \delta(s-m^2_\H) ;
\label{d8}
\end{eqnarray}
Note that the PV method gives $\int_{-a}^b x^{-2} dx = -a^{-1} - b^{-1}$, even
when the double pole is straddled by $-a$ and $b$.  The SDF method attempts to
isolate the off-shell part by subtracting what the scattering rate ``would have
been'' if the resonance were infinitesimally narrow.  Although these
alternatives at first look quite different from each other, by writing
$\pi\delta(s-m^2_\H)$ as $\epsilon/((s-m_\H^2)^2+\epsilon^2)$ one can show that
they are actually equal to each other in the limit as $\Gamma\to 0$; writing $x
= (s-m^2_\H)$, and $\epsilon'=\epsilon\Gamma$, the difference is
\beq
        {x^2-\epsilon'^2\over(x^2+\epsilon'^2)^2}{\rm\ versus\ }
        {x^2-\epsilon'^2\over(x^2+\epsilon'^2)^2+ x^2\Gamma^2}.
\label{d8a}
\eeq
In fact this is as good a correspondence as one should hope for since the whole
concept of asymptotic states which have thermal distributions breaks down if
they become very broad resonances.  Thus two subtraction schemes should be
considered equally good if they differ only by terms of order $\Gamma^2$. We
compare the PV and SDF methods in Table 3, of which the former was used in the
top-quark contributions shown in Table 2.

\begin{table}
\def\t{\times}
\def\p{\phantom{-}}
\begin{center}
\begin{tabular}{|c|c|c|} \hline \hline
  $m_\H/T$   &       PV              &    SDF             \\ \hline
      0.4    &     $\p 1.52\t 10^{-5}$  &   $\p 1.52\t 10^{-5}$ \\
      0.8    &     $\p 1.43\t 10^{-5}$  &   $\p 1.42\t 10^{-5}$ \\
      1.2    &     $\p 1.04\t 10^{-5}$  &   $\p 1.04\t 10^{-5}$ \\
      1.6    &     $\p 5.24\t 10^{-6}$  &   $\p 5.33\t 10^{-6}$ \\
      2.0    &     $\p 3.95\t 10^{-7}$  &   $\p 3.58\t 10^{-7}$ \\
      2.4    &     $-3.53\t 10^{-6}$    &   $-3.53\t 10^{-6}$   \\
      2.8    &     $-6.15\t 10^{-6}$    &   $-6.16\t 10^{-6}$   \\
      3.2    &     $-7.14\t 10^{-6}$    &   $-7.55\t 10^{-6}$   \\
      3.6    &     $-7.70\t 10^{-6}$    &   $-7.67\t 10^{-6}$   \\
\hline\hline
\end{tabular}
\end{center}
\noindent Table 3: Contributions to $I_{top}/K$ (see table 2) due to
$s$-channel Higgs boson scattering.  We compare the principal value
prescription used in table 2 to the delta function subtraction method.
\end{table}

The negative values for $m_\H/T>2.0$ may be alarming at first, since they are
supposed to be the off-shell part of a scattering rate.  Although our results
are completely insensitive to the negative contributions because of their
relative smallness compared to the decay rate, it is interesting to speculate
on their meaning.  We take them as being suggestive that the decay rate $I_D$
gives an overestimate of the Higgs-mediated processes at large values of
$m_\H/T$.  In this regime, due to strong couplings to itself or the top quark,
the Higgs boson is starting to become a rather broad resonance where it may not
be sensible to consider the decays; rather one should include the scatterings
only, without subtracting the resonance from the $s$-channel. These give a
smaller contribution than the decays because the area under the resonance curve
gets diminished by the Boltzmann suppression in the function $K_1(\sqrt{s}/T)
\sim \exp(-\sqrt{s}/T)$ (eq.~(\ref{d6})).  But since the off-shell scatterings
are completely negligible compared to the decays in the large $m_\H/T$ regime,
the question of whether to take the negative cross-section seriously is
moot--its contribution is negligible.

\subsection{$\Delta L=2$ Scatterings}

In a later section we will also be interested in how $\Delta L=2$ lepton number
violating interactions come into equilibrium, in connection with baryon number
erasure.  The operator we consider, which gives rise to neutrino masses,
has dimension five:
\beq
   \sum_{ij}{m_{ij}\over 2v^2}(\bar L_i H)(1+\gamma_5)(H^T L_j^c) +{\rm h.c.},
\label{d9}
\eeq
where $m_{ij}$ is the neutrino mass matrix in the flavor basis, $v$ is the
246 GeV VEV of the Higgs field $H$, and $L_i$ are the lepton doublets.

The sum over channels of the cross sections for the processes $\el\nu_\ell\to
h^*h^-$ and $\el\ell_i\to h^- h^-$ must be properly weighted by the factors
$S_\ell$ shown in the Boltzmann equation (\ref{BE01b}) for $\el$, because of
the complication of having identical particles in some of the initial or final
states. We find that
\beq
        \sum_c \sigma_{c,\ell}(s) = {1\over 4\pi v^4}\left\{
	\begin{array}{ll}
        5|m_{ee}|^2, & \ell=e \\
        6|m_{e\ell}|^2, & \ell=\mu,\tau.
	\end{array} \right.
\label{d10}
\eeq
This is what we must substitute into eq.~(\ref{d6}) to obtain the quantities
$I_{\Delta L,\ell}$, which appear in eq.~(\ref{d3}) for the rate of lepton
violation. The various channels respectively contribute the relative weights
$(2,2,4,1,1)|m_{ee}|^2$ for $\el\el\to h^-h^-$, $\el\nu_e\to h^-h^*$, $\el
h^+\to\el^c h^-$, $\el h^+\to{\nu_e}^c h^*$, $\el h\to{\nu_e}^c h^-$ and
$(4,2,4,1,1)|m_{e\ell}|^2$ for $\el\ell_\L\to h^-h^-$, $\el\nu_\ell\to h^-h^*$,
$\el h^+\to{\ell_\L}^c h^-$, $\el h^+\to{\nu_\ell}^c h^*$,$\el h\to{\nu_\ell}^c
h^-$ with $\ell=\mu,\tau$.  In the approximation of ignoring masses compared to
$s$, the integral in eq.~(\ref{d6}) can be done exactly,
\beq
        I_{\Delta L,\ell} = {T^6\over\pi^4}\sum_c\sigma_{c,\ell}.
\label{d11}
\eeq
This simplification occurs because of the constant cross section.  Of course
at very high temperatures of order the mass $M$ of the heavy neutrino which
must have induced the operator (\ref{d9}), the cross section will exhibit a
form factor behaving as $M^2/(T^2+M^2)$, coming from the propagator.  Thus we
are assuming that $M$ is greater than the temperatures of interest when we use
eq.~(\ref{d11}).  It will be shown later that as long as $M$ exceeds 10 TeV,
our neglect of the propagator is justified.

\def\lsim{\;\raise0.3ex\hbox{$<$\kern-0.75em\raise-1.1ex\hbox{$\sim$}}\;}
\def\gsim{\;\raise0.3ex\hbox{$>$\kern-0.75em\raise-1.1ex\hbox{$\sim$}}\;}

\section{The case of conserved $B - L$}

The importance of the weak electron chirality-changing interactons
in the context of preserving the baryon asymmetry was first discussed by the
authors of ref.\ \cite{cdeo3} (CDEO). They considered whether in a universe
with primordial $B-L=0$, the right-handed electrons could stay out of
equilibrium long enough so that the primordial baryon asymmetry, kept nonzero
by the asymmetry in the $e_R$ species, does not get washed out completely
before the sphaleron interactions fall out of equilibrium.  While their result
was negative for the preservation of the baryon asymmetry, they noted that the
outcome of the computation is exponentially sensitive to the strength of the
chirality-changing interactions. Therefore even a rather small change in the
interaction rates could alter the conclusion.

Indeed, the derivation of CDEO contained a number of inaccuracies with
the potential to change the outcome significantly. In particular, their
neglect of the final state particle masses leads to an overestimate of
the rate of Higgs decays, and therefore to an underestimate of the final
asymmetry.  As we have seen in the section 3 above, the finite
mass effects are very important for small (thermal) Higgs boson
masses. On the other hand, CDEO neglected all scattering processes,
which we found to be of particular importantance when the
decay process is suppressed by the phase space effects.
Armed with our accurate expressions for all tree level chirality
changing rates computed in the previous section, as well as with the
carefully formulated evolution equations derived in section 2, we
can now re-examine and answer definitively the question CDEO posed.

As explained in section 3, the evolution equations for the asymmetries depend
on the  particular set of equilibrium conditions and conservation laws
characterizing the system. Here the set of boundary conditions is given by the
equilibrium equations (\ref{chem}--\ref{charge}) supplemented with the
additional constraint of vanishing $B-L$. In fact we only assume that
primordially $(B-L)_p = 0$; the vanishing of $B-L$ at any lower temperature is
guaranteed by the first three of the conservation laws (\ref{claws}), which
also imply the conservation of $B-L$. In terms of the chemical potentials we
then find the constraint
\begin{equation} B-L = 12\mu_{u_L} - 3\sum_\ell
\mu_{\ell_L} - 2\mu_{e_L} - \mu_{e_R} + 2\mu_0 = 0.  \label{BminusL}
\end{equation}
{}From eqs.~(\ref{SphEQ}--\ref{charge}) and (\ref{BminusL}) one can readily
find that
\begin{eqnarray}  \mu_0 &=& \frac{5}{153}(\mu_{e_R} - \mu_{e_L})
\nonumber \\ \mu_B &=& \frac{44}{153}(\mu_{e_R} - \mu_{e_L}).
\label{mu0andB1}
\end{eqnarray}
Note that if we further assumed the $e_R$ population is in equilibrium,
i.e. $\mu_{e_R}-\mu_{e_L}=-\mu_0$, then only the trivial solution would
exist: $\mu_0 = B = L = 0$, in agreement with the standard analysis of
ref.~\cite{ht}.

Our goal is to follow the evolution of the $\er$ asymmetry using
eq.~(\ref{BEM}), where from (\ref{mu0andB1}) we see that the constants $a,b$
are given by $a = -b = 5/153$.  However, we need not consider coupled equations
for $L_\el$ and $L_\er$ because the latter is determined through the
conservation law $\frac{1}{3}B-L_e = C_{e,p}$.  Eliminating $B$ from this
equation using (\ref{mu0andB1}) for $\mu_B \propto B$, we find that
\begin{equation}
L_{e_L} =  -\frac{415}{962}L_\er - \frac{459}{962}C_{e,p}.
\label{Labove}
\end{equation}
Inserting this into eq.~(\ref{BEM}) then gives the simple first order equation
for $L_\er$
\begin{equation}
\frac {\rd L_{e_R}}{\rd t} = - {\Gamma} (L_{e_R}+ \frac{1}{3}C_{e,p}).
\label{BEa}
\end{equation}
where
\begin{equation}
{\Gamma} =  \frac{711}{481}\Gamma_{RL}
\cong {17.7\over T^3}(I_D + I_G + I_{top}).
\label{Gam}
\end{equation}
The quantities $I_x/T^3$ are the thermally averaged rates computed in section
2. Eq.~(\ref{BEa}) is easily integrated to yield a general solution for
$L_{e_R}$ as a function of time that is completely specified in terms of
primordial quantities:
\begin{equation}
L_{e_R}(t) =  - \frac{1}{3}C_{e,p} + (L_{e_R,p} + \frac{1}{3}C_{e,p})
e^{-\int_{t_0}^t \rd t' \Gamma}.
\label{BEq}
\end{equation}

Interestingly, the right handed electron asymmetry does not tend to zero, but
instead relaxes to a fixed value $-\frac{1}{3}C_{e,p}$  which may be even
larger in absolute value than the primordial value of $L_{e_R}$. This is
contrary to the assertion of CDEO that $L_{e_R} \rightarrow 0$. What happens is
that at the fixed point the right and left chiral asymmetries become equal
under the additional constraint of $(\frac{1}{3}B-L_e)$-conservation. This
results in erasing the baryon and lepton asymmetries, $B,L\to 0$, as can be
inferred from eq.~\ref{mu0andB1}. It is useful to write down an expression for
$B$ directly as a function of time
\begin{equation}
B(t) = B_{eq} e^{-\int_{t_0}^t
\rd t' \Gamma},
\label{Basfoft}
\end{equation}
where the equilibrium value of $B$ much above $T_*$ is given by
\begin{equation}
B_{eq} = \frac{66}{481}(C_{e,p} + 3L_{e_R,p}).
\label{Bequil}
\end{equation}
As expected, the value of $B$ decreases exponentially in time, simultaneously
with the
equilibration of the $e_R$ population.  It is conceivable that the
initial value for $B_{eq}$ could be as large as $B^{eq} \sim 10^{-2}$ in some
grand unified theories, although typically it will be suppressed by small
coupling constants and $CP$-violating phases.  We will allow $B^{eq}$ to vary
in what follows.

Equation (\ref{Basfoft}) is our main result in this section.
Although we will compute the rate $\Gamma$ and its time integral for
(\ref{Basfoft}) numerically, to get a feeling for the relevant
scales it is useful to find an analytic approximation for $B_{\rm
final}$.
We can do this by employing linear fits to the values of
the functions $I_x$ given in table 2, and integrating the ensuing
fit functions analytically. Using a simple linear regression formula, we
find
\beq
\Gamma_{RL} \simeq  3.8\times 10^{-3} T h_e^2 f(m_\H (T)/T),
\label{gammafit}
\eeq
where the fit function
\beq
f(x) \equiv  (-1.1+3.0x) + 1.0 + h_t^2 (0.6-0.1x)
\label{fx}
\eeq
has been normalized so that the middle term, due to gauge/Higgs scattering, is
unity. Because the number of degrees of freedom is constant, $g_* = 106.75$,
entropy is conserved and we can use the approximation $\dot{T}/T = -H$. Then,
neglecting the threshold effect in the decays, we obtain the result
\begin{equation}
\int_0^{t_c} \rd t \Gamma \simeq  350 \left(\frac{100\,{\rm GeV}}{T_c}\right)
\left[ (-1.1+2.4x_H) + 1.0 + h_t^2 (0.6-0.09x_H) \right],
\label{analytic}
\end{equation}
where we have defined the asymptotic high-temperature limit of the
Higgs boson thermal mass as
\beq
	\lim_{T\to\infty} m_\H(T)/T = \sqrt{2D} \equiv x_H
\label{xh}
\eeq
The approximation (\ref{analytic}) treats the decay part too roughly for small
thermal Higgs masses, and therefore cannot be used to obtain figure 4a. However
it is a very good estimate for the total integrated rate over the whole
interesting range of $x_H\la 2$. It should be noted that for a large range of
thermal Higgs masses, $x_H \la 0.9$, the scattering contribution to the
integrated rate dominates over the decay part.

In figure 5 we show the result of our computation for the final baryon
asymmetry in units of the initial asymmetry (\ref{Bequil}), as a function of
the physical Higgs and top quark masses. For comparision to the work of CDEO,
we show in figure 5a what the outcome would have been, had we included only the
corrected decay contribution to the total rate. The curves are contours of
constant $B_{\rm final}$. The initial upward bending of the curves as a
function of the zero-temperature Higgs boson mass $m_0$ occurs because the
endpoint of the integration, the electroweak phase transition temperature
$T_c$, increases with $m_0$. The resulting decrease in the integrated rate can
be compensated by increasing $m_t$. As $m_0$ continues to increase, the thermal
Higgs mass starts to show its dependence on $m_0$ (cf.\ eq.\ \ref{d3}) leading
to a significant increase in the rate as a function of $m_0$. Simultaneously
the phase transition temperature reaches a plateau and therefore the contours
turn down at large values of $m_0$.

It is noteworthy, and in contrast to the result of CDEO, that their mechanism
could have appeared to be successful in protecting the baryon asymmetry for a
small region of experimentally allowed parameter space, had they considered
only the more exact expression for the Higgs boson decay rate. The main reason
for this would-be positive result is the suppression of the decay rate for
small values of $x_H$, where the phase space volume is small. When the
scattering contributions are included however, the conclusion changes
drastically. We show the result of the numerical integration in figure 5b. In
particular, because of the gauge scatterings the final baryon asymmetry is
completely negligible over the entire parameter space. Thus the scatterings
play a decisive role in the most relevant region of parameters for the problem.
 In the regime where the decay contribution dominates over the scattering rate,
$B_{\rm final}$ is even more suppressed than in the region of parameters shown
in figure 5b, which extends only up to $x_H \sim 0.55$.

Our conclusion is qualitatively the same as that of CDEO: it is not possible to
protect the baryon asymmetry of a $B-L=0$ universe by temporarily storing it in
the asymmetry of the $e_R$ species. However, the way in which we reached this
conclusion differed from theirs in important respects and moreover, the
robustness of our result erases any doubts expressed by CDEO, that some subtle
thermal effects might alter the outcome.

\def\H{{\sss H}}
\def\G{{\sss G}}
\def\L{{\sss L}}
\def\R{{\sss R}}
\def\nmp{\langle m_{ei}\rangle^2}

\section{Examples when $B-L$ is not conserved}

Although we have seen that an $\er$ asymmetry cannot save the baryon asymmetry
from sphaleron effects when $B-L=0$,  it can have a dramatic effect when
$B-L\ne 0$, as one might have in SO(10) grand unified theories where $B-L$ gets
broken at the GUT scale.  In this section we will explore the consequences of
an $\er$ asymmetry in conjunction with sphalerons and other $B$- and
$L$-violating interactions that hitherto were believed to completely erase $B$.

It is expected that any global symmetries in nature are only approximate;
they might be broken by exotic gravitational means such as microscopic black
holes or wormholes, or more mundane reasons like the fact that one can
construct renormalizable $B$- and $L$-violating interactions in supersymmetry
using the squark and slepton fields.  Such effects may be described by
nonrenormalizable operators in the low energy effective theory.  Assuming that
a
primordial baryon asymmetry survives allows us to constrain the strength
of these new interactions. These constraints were first applied to
the $\Delta L = 2$ operator (\ref{d9}) by Fukugita
and Yanagida \cite{fy2}. They were generalized to other operators
in refs.\ \cite{cdeo12,sonia}.

Quite generally, an operator of dimension $4+n$ would have a dimensionful
coefficient of the form $M^{-n}$, and would mediate a process at high
temperatures whose rate scales like $\Gamma \sim T^{2n+1}M^{-2n}$.  This rate
is to be compared to the Hubble expansion rate $\sim T^2/M_{Pl}$.  Insisting
that the interaction be out of equilibrium below the temperature $T_0$ at
which sphalerons came into equilibrium ($10^{12}$ GeV), or when the baryon
asymmetry was formed, whichever was smaller, leads to a lower bound on the
scale of new physics:
\beq
        M \gsim T_0\left({M_{Pl}\over T_0}\right)^{1/2n}.
\label{e1}
\eeq
In supersymmetric models it was argued \cite{iq} that due to additional
anomalies which can temporarily protect the asymmetry (until the effects of
supersymmetry breaking kick in), the maximum temperature should be at $T_0 \sim
10^8$ GeV rather than $T_0 \sim 10^{12}$ GeV. Interestingly, in the context of
inflation the maximum temperature should not surpass the thermalization
temperature \cite{eeno}, which could be as low as $\sim 10^5$ GeV \cite{cdo}.
We will argue that actually none of these values of $T_0$ are the correct
temperature to use in setting this bound whenever an $\er$ asymmetry is
present, as it will prevent the baryon asymmetry from relaxing to zero,
essentially because of the charge neutrality of the universe. Rather, one
should use the temperature $T_*$ at which $\er$-violation comes into
equilibrium and allows the $\er$ asymmetry to disappear.  In this section we
will compute $T_*\simeq 1$ TeV.  Hence constraints like (\ref{e1}) might be
weakened by as many as nine orders of magnitude, by standard model physics
which has been known for well over a decade! Note, however, that our arguments
here do not affect the limits imposed on renormalizable  $B$- or $L$-violating
operators. For these, the strongest bound always comes from applying the limit
at $T_c$ (when sphalerons disappear) \cite{cdeo12}.

Before computing $T_*$, we will follow the evolution of the baryon and $\er$
asymmetries more exactly, for a particular example of a nonrenormalizable
operator, that which would be responsible for the electron neutrino mass.
An example employing explicit $B$-violation will also be given.

\subsection{Neutrino masses}

The best known example of $B-L$ violation is the possibility of neutrino
Majorana masses, which would be generated by the dimension five operator,
$\sum m_{ij}v^{-2}(L_iH)(L_jH)$.  To avoid an erasure of the baryon asymmetry,
upper limits on $m_{ij}$, the neutrino mass matrix elements, must be satisfied
\cite{fy2,ht,nb}.  The flavor-dependence of the operator will be important
since any approximately conservered quantities (such as $\frac{1}{3}B - L_i$)
would protect the asymmetry \cite{nb,dr}.  If for all flavors the operator was
so weak as never to have been in equilibrium, and $B - L = 0$, then as shown in
the previous section, the baryon asymmetry would be erased, up to small mass
effects \cite{krs2,dr}.  If, on the other hand, one or two of the lepton
generations were violated by this operator and the remaining generations were
out of equilibrium, then even if $B = L = 0$ (or more generally, $B-L = 0$) a
baryon number would be generated provided some initial lepton flavor asymmetry
existed \cite{cdeo3}.   But if all three flavors were violated down to low
enough temperatures, the baryon asymmetry would be erased.  How low is low
enough? The answer, which we will denote by $T_*$, depends crucially on the
decoupling temperature of the chirality-flipping electron interactions.  In
this section we will determine $T_*$, as well as the size of the neutrino mass
matrix elements which would be compatible with preserving a primordial baryon
asymmetry.

Supposing that the masses of $\nu_\mu$ and $\nu_\tau$ were
sufficiently large (several keV) that the separate lepton flavors $L_\mu$ and
$L_\tau$ were strongly violated during the sphaleron epoch, we would insist
that $L_e$ is not similarly violated, if we want to keep sphalerons and
neutrino
mass effects from erasing the baryon asymmetry.   Since we are working above
the temperature at which the Higgs gets its VEV, however, the effect of
interest
will be scatterings of leptons into bosons.

By assumption, the $\Delta L=2$ scattering processes $L_i L_j\to HH$ will be in
equilibrium if $i$ and $j$ are either $\mu$ or $\tau$, which means that in
place of the conservation laws for $\frac{1}{3}B- L_\mu$ and
$\frac{1}{3}B-L_\tau$ in eq.~(\ref{claws}), we will have two nonstandard
equilibrium conditions
\beq
        \mu_{\mu_L} = \mu_{\tau_L} = - \mu_0
\label{e3}
\eeq
involving the Higgs chemical potential.  As in the previous section, the new
constraints allow us to solve for the parameters $a$ and $b$ in
eq.~(\ref{muHiggs}). We find that
\begin{eqnarray}
        \mu_0 &=& \frac{3}{55}\mu_{e_R} + \frac{1}{11}\mu_{e_L} \nonumber \\
        \mu_B &=& \frac{24}{165} \mu_{e_R} - \frac{12}{11}\mu_{e_L}.
\label{e4}
\end{eqnarray}

In addition to the equilibrium conditions (\ref{e3}) for the mu and tau
species, the new interaction provides a source of electron-type lepton number
violation which goes out of equilibrium at a temperature depending on the size
of the neutrino mass matrix elements $m_{ei}$.  The effect is parametrized by
the integrals $I_{\Delta L,i}$ in the Boltzmann equation (\ref{BEM}),
calculated in section 3. The evolution equations for the two electron
chiralities then become
\begin{eqnarray}
        \frac {\rd L_{e_R}}{\rd t} = &-& \Gamma_{RL}
        \left(\frac{58}{55}L_{e_R}-\frac{10}{11}L_{e_L}\right) \nonumber \\
        \frac {\rd L_{e_L}}{\rd t} = &-&
        \Gamma_{\Delta L}\left(\frac{24}{13}L_{e_L}+\frac{6}{65}L_{e_R}\right)
        - \frac{157}{390} \frac{\rd L_{e_R}}{\rd t},
\label{e5}
\end{eqnarray}
where $\Gamma_{RL}$ was defined in (\ref{Gamma}) and\footnote{The relative
coefficient of $|m_{ee}|^2$, which was wrong in ref.~\cite{cko2}, has been
corrected here.}
\begin{eqnarray}
	\Gamma_{\Delta L} &=& \Gamma_{\Delta L,e} +
	\frac{1}{2}\Gamma_{\Delta L,\mu} +
	\frac{1}{2}\Gamma_{\Delta L,\tau}\nonumber\\
	&=& \frac{3T^3}{\pi^5 v^4}(5|m_{ee}|^2 +3|m_{e\mu}|^2 + 3|m_{e\tau}|^2).
\label{e6}
\end{eqnarray}
Using the results of section 3, we have integrated these equations starting at
high temperatures $T\gg 100$ TeV where the $\er$-violating interactions have
not yet reached equilibrium.  We consider a range of initial conditions for
$L_\er$, including the largest reasonably imaginable asymmetry in $\er$, which
would occur if there were only $\er^-$ and no $\er^+$; then dividing $n_\er$ by
the total entropy gives $L_\er(0) \cong 10^{-2}$.  The initial value for
$L_\el$ is given by solving the chemical potential equations subject to the
constraint that the $L_e$-violating interactions {\it are} in equilibrium:
$\mu_\el = -\mu_0$, as in eq.~(\ref{e3}).  We find that $L_\el =
-\frac{1}{20}L_\er$ and $B = \frac{1}{5}L_\er$ initially (see section 5.1,
below).  The resulting value of the baryon asymmetry is
then determined by the final lepton asymmetries using eq.~(\ref{e4}): $B =
\frac{24}{165}L_\er - \frac{12}{11}L_\el$.

In figures 6a and 6b we show the evolution of the asymmetries $L_{e_R}$,
$L_{e_L}$ and $B$ as a function of the temperature for $m_t=m_H = 100$ GeV. In
figure 6a we have assumed that the electron lepton number violation has gone
out of equilibrium before the $\er$-$\el$ transitions come into equilibrium.
Initially all asymmetries have their high temperature equilibrium values. As
the right handed electrons come into equilibrium, the asymmetries relax to
their
low-temperature equilibrium values, which are of the same order as the
asymmetries above the transition region and are calculable in terms of the
primordial $L_{e_R}$ asymmetry.  In figure 6b we assume that the neutrino mass
parameter
\begin{equation}
	\nmp\equiv \frac{5}{3}|m_{ee}|^2 + |m_{e\mu}|^2 + |m_{e\tau}|^2
\label{NMP}
\end{equation}
takes the value $\langle m_{ei} \rangle = 5$ keV.  Now the asymmetries fall
exponentially within the region where the lepton number violation and the
$\er$-$\el$ transitions (with rate $\Gamma_{e_R\leftrightarrow e_L}$) are both
in equilibrium. A new equilibrium is reached above the electroweak phase
transition temperature, $T_c \simeq 154$ GeV, as the lepton number violating
interactions fall out of equilibrium. The lesson to be learned from figure 6b
is that, in order to erase the asymmetries, the temperature must fall somewhat
below that at which $\Gamma_{e_R\leftrightarrow e_L}/H=1$, as is often assumed
in simple estimates.

To obtain a conservative bound on $\langle m_{ei} \rangle$, we must assume that
the lepton number violation is strong enough to decrease the largest imaginable
initial baryon asymmetry down to the smallest value of
the present baryon asymmetry $B_{\rm min} \simeq 4\times 10^{-11}$ inferred
from big bang nucleosynthesis
\cite{Keith}.  For the former value we take $B_{\rm initial}=2\times 10^{-3}$,
corresponding to $L_{e_R,p}=10^{-2}$.  Moreover, we will assume the least
favorable values $m_t=90$ GeV and $m_H=60$ GeV for the unknown masses of the
top quark and the Higgs boson. In this way one obtains the bound
\begin{equation}
\langle m_{ei} \rangle \la 20 {\rm\ keV},
\label{finalbound}
\end{equation}
an order of magnitude less stringent than the rough estimate given in our
letter \cite{cko2}. This result can also be inferred from figure 7, where we
show the results of a more complete computation of the bound as a function of
the initial asymmetry.  We also display the dispersion in the bound that
results from the above-mentioned  uncertainty in the present baryon asymmetry
and the top quark and Higgs boson masses.  The shaded area corresponds to the
intervals 90 GeV $<m_t<$ 300 GeV, 60 GeV $<m_\H<$ 1 TeV and $4\times 10^{-11}
<B_{\rm final}<6\times 10^{-11}$.

A simple analytic estimate for the bound (\ref{finalbound}) can be obtained by
requiring that the rate of lepton number violation, or more generally, any
other baryon or lepton number violation under consideration, becomes comparable
to the expansion rate after the time $t_*$ defined by
\begin{equation}
\int_{0}^{t_*} {\rm d} t \; \Gamma_{LR} = \ln(B_{\rm initial}/4\times10^{-11})
\simeq 18,
\label{simplebound}
\end{equation}
using our assumption that $B_{\rm initial} = 2\times 10^{-3}$. To solve
equation (\ref{simplebound}) for the corresponding temperature $T_*$, we use
the linear fit for $\Gamma_{LR}$ derived in section 4.  After analytically
integrating the fit function and expanding in the small parameter $T_c/T_*$ we
find that
\begin{equation}
T_* \simeq 1.3 f(x_H)  \rm \; TeV,
%\cdot (1.0 + (-1.1+3.0x_H))
\label{Tstar}
\end{equation}
where $f(x_H)$ is as in eq.~(\ref{analytic}). Again assuming that $m_t=90$ GeV
and $m_H=60$ GeV, we get $x_H\simeq 0.41$ and then $T_* \simeq 1$ TeV.  Note
that while we have so far loosely spoken of $T_*$ as the temperature at which
$\er$-$\el$ transitions come into equilibrium, it is actually given by the
considerably smaller temperature (\ref{Tstar}), which is relevant for obtaining
the bounds on various baryon or lepton number violating operators. Let us
finally note that using (\ref{Tstar}) in the rough condition for freezeout of
the lepton-violating interactions,
\beq
	\left.{\Gamma_{\Delta L} \over H}\right|_{T=T_*} = 1,
\label{freezeout}
\eeq
gives a bound $\langle m_{ei}\rangle \la 21$ keV, in good agreement with the
accurate numerical result (\ref{finalbound}).

\subsection{Baryon number violating operators}

Apart from neutrino masses, there is a host of possible $(B-L)$-violating
interactions that have explicit baryon number violation.  To further illustrate
the relevance of right-handed electrons for preserving a baryon asymmetry, we
will show what happens when several such operators are in equilibrium during
some epoch.

It is convenient to rewrite the sphaleron and electric charge constraints on
the chemical potentials in the form
\begin{eqnarray}
        \frac{35}{3}H + \frac{1}{2}B &=& \frac{1}{2}L_e
        +\frac{2}{3}(L_\mu+L_\tau) + \frac{1}{2}L_\er;\nonumber \\
        \frac{2}{3}H + \frac{3}{4}B &=& -\frac{1}{2}L_e
        -\frac{1}{3}(L_\mu+L_\tau) + \frac{1}{2}L_\er,
\label{e7}
\end{eqnarray}
where $H$ is the Higgs boson asymmetry, $L_e\equiv 2L_\el+L_\er$ is the
asymmetry for total electron-type lepton number, and similarly for the other
flavors.  We have noted that in the minimal standard model, there are three
conserved quantities $\frac{1}{3}B-L_i$ and one which is approximately
conserved at high temperatures, $L_\er$.  In the absence of new interactions we
would solve for the baryon and Higgs chemical potentials in terms of the
conserved quantities.

When sufficiently fast interactions are introduced which violate any of the
conservation laws, eqs.~(\ref{e7}) must be supplemented by the new chemical
equilibrium conditions associated with these interactions.  Of course we do not
want to break all four of the symmetries, since then the solution to the system
of equations would be the vanishing of all asymmetries.  We do wish to break
the particular linear combination $B-L$, however, since only in this case
do we get anything new from our observations concerning right-handed
electrons.  If $B-L$ is conserved during the epoch when sphalerons are in
equilibrium, yet has a nonzero value, this by itself is enough to insure a
nonvanishing baryon asymmetry, regardless of the electron \cite{ht}.

For example, consider the $\Delta B = 2$ operator
\beq
	M^{-5}(\bar u_{\R} d^c_{\R}) (\bar d_{\R} d^c_{\R})
	(\bar u_{\R} d^c_{\R}).
\label{e8}
\eeq
It conserves the linear combination
\beq
	2L_e - L_\mu - L_\tau \equiv C.
\label{e9}
\eeq
(Although it conserves both $L_e-L_\mu$ and $L_e-L_\tau$ separately, only the
sum of these is relevant since $L_\mu$ and $L_\tau$ enter eqs.~(\ref{e7}) in
the combination $L_\mu+L_\tau$.)  It also leads to the equilibrium constraint
\beq
	2\mu_{u_\R} + 4\mu_{d_R} = 0 \quad \Rightarrow \quad B = 4H,
\label{e10}
\eeq
where we used eq.~(\ref{chem}) to eliminate the chemical potentials of the
right-handed quarks in favor of $\mu_{q_\L}$ and the relation $\mu_B =
12\mu_{q_\L}$.  It is now possible to solve for $B$ and the total lepton number
$L$ using eqs.~(\ref{e7},\ref{e9}-\ref{e10}), and in particular their
difference\footnote{This is similar to the result in  ref.~\cite{cko2}
but written in terms of the conserved quantities $C$ and $L_\er$.}
\beq
	B-L = \frac{3}{68}(C - 9L_{\er,p}),
\label{e11}
\eeq
where we subscripted $L_\er$ to emphasize that it retains its primordial value
until $T_*$, the temperature at which the rate of $\er$-violation starts to
exceed the expansion rate of the universe.  Let us suppose that the new
interaction (\ref{e8}) goes out of equilibrium at some temperature $T'$ below
the baryogenesis scale but above $T_*$. Then for $T'> T > T_*$, the asymmetry
$B-L$ maintains its equilibrium value (\ref{e11}), since nothing occurs that
would change it.  As the temperature falls below $T_*$, $B-L$ continues to be
conserved, even though $L_\er$ itself changes.   At this point, the complete
equilibrium conditions assumed by Harvey and Turner prevail, and we can use
their results to determine the final $B$ in terms of $B-L$.

This result is remarkable in two respects.  First, $B-L$ in (\ref{e11}) is
completely independent of its initial value, at temperatures before the $\Delta
B=2$ operator came into equilibrium.  One possibility is that it was initially
zero, in which case the new operator actually worked in conjunction with
sphalerons to {\it create} the baryon asymmetry, which would have been zero in
its absence.  So we have here a new way to convert a primordial lepton
asymmetry into baryons. Secondly, even though this mechanism depends on the
conservation of $\er$ number, it does not require that $L_\er$ be nonzero.  It
was enough to have an asymmetry between the different flavors of leptons in
this example.

In general we can violate all three of the quantities $\frac{1}{3}B-L_i$
and still obtain a nontrivial solution for $B$.  For our second example,
we take the operators
\beq
	M^{-5}(\bar u_\R {u_\R}^c)(\bar u_\R{\ell_{i,\R}}^c)(\bar L_j \tau_2
	{L_k}^c),
\label{e12}
\eeq
where the indices $i,j,k$ are generational and $\tau_2$ is the Pauli matrix.
Although there are numerous possible combinations of $i,j,k$, consider at first
any three which break all three symmetries.  These give relations among the
chemical potentials which as before can be expressed entirely in terms of
left-handed fields using the constraints (\ref{chem}) from Higgs boson
interactions to eliminate the right-handed fields.  After so doing, the new
relations are
\beq
        3\mu_{q_\L} = -\mu_{\mu_\L} = -\mu_{\tau_L} = \mu_0,
\label{e13a}
\eeq
which translate directly to
\beq
        B = -L_\mu = -L_\tau = 4H
\label{e13b}
\eeq
in terms of the asymmetries.  We can solve these together with eqs.~(\ref{e7})
for all of the asymmetries in terms of $L_\er$:
\begin{eqnarray}
        H &=& \frac{1}{20}L_\er; \nonumber\\
        B &=& -L_\mu = -L_\tau = \frac{1}{5}L_\er; \nonumber\\
        L_e &=& \frac{9}{10}L_\er.
\label{e14}
\end{eqnarray}
The final baryon asymmetry will again be determined by $B-L$, whose equilibrium
value above $T_*$ is
\beq
	B-L = -\frac{3}{10}L_{\er,p}.
\label{e15}
\eeq
as first shown in \cite{cko2}.

If the primordial asymmetry in $\er$ was sufficiently large,  the new $B$- and
$L$-violating interactions (\ref{e8},\ref{e12}) were harmless for the baryon
asymmetry, so long as they fell out of equilibrium above $T_*$.  As outlined in
the introduction to this section, this leads to upper bounds on the masses in
their coefficients \cite{cdeo12}.  Let us  examine the bound for the dimension
nine operators $M^{-5}qqqqqq$ and $M^{-5}qqqlll$ we have just discussed.  For a
particular scattering channel, say for a $3\to 3$ process, we saw in section 2
that the relevant rate is typically given by $\Gamma \sim 12I/T^3$, where
\beq
	I \cong \int \prod_{i=i}^6\rd\Pi_i(2\pi)^4
	\delta(p_1+p_2+p_3-p_4-p_5-p_6) |{\cal M}|^2 e^{-\beta(E_1+E_2+E_3)}
\label{e16a}
\eeq
and we can estimate the matrix element as
\begin{eqnarray}
	|{\cal M}|^2 &=& 8M^{-10}p_1\ddot p_2\, p_3\ddot p_4\, p_5\ddot p_6
		\nonumber\\
	&\cong& 8M^{-10}\prod_i E_i.
\label{e16b}
\end{eqnarray}
There are $6\ddot 5\ddot 4$ different channels for the $3\to 3$ processes and
$6\ddot 4$ for the $2\to 4$ processes.  Putting everything together and
evaluating the integrals, we find a rate of
\beq
	\Gamma \cong 288 (2/\pi)^{14} T \left({T\over M}\right)^{10},
\label{e16c}
\eeq
which, if we insist it be less than the Hubble rate at temperature $T_0$, gives
the lower limit on $M$ of
\beq
	M > 0.7 T_0 \left({M_{\rm Pl}\over T_0}\right)^{1/10},
\label{e16d}
\eeq
in remarkably good agreement with the dimensional analysis estimate (\ref{e1}).
Previous limits on this operator \cite{cdeo12} ranged from $10^3 - 10^{12}$ GeV
for $T_0$ between $10^2 - 10^{12}$ GeV.  We have shown that the appropriate
temperature is $T_0 = T_*\sim 1$ TeV.  The difference for the bound on the
scale of new physics is marked:
\beq
        M > \left\{
	\begin{array}{ll}
        4\times 10^{12}\ {\rm GeV}, & T_0 = 10^{12}\ {\rm GeV} \\
        30\ {\rm TeV}, & T_0 = T_*
	\end{array} \right..
\label{e16e}
\eeq
and hence the limit is much closer to the weaker of the bounds given in
\cite{cdeo12}. Limits on other nonrenormalizable operators are similarly
weakened.

The reader may wonder how general our results are.  Did we need exactly three
of the operators (\ref{e12}) in equilibrium, and did it matter which three?
Intuitively we expect that the answer to both questions is no as long as all
three symmetries $\frac{1}{3}B-L_i$ are broken, since the only thing keeping
the chemical potentials from vanishing is the necessity of maintaining charge
neutrality of the universe by canceling the charge of the conserved $\er$
asymmetry. Therefore it should not matter how many or which kind of operators
do the job of breaking $\frac{1}{3}B-L_i$.

It is easy to prove that having other operators in addition to all those we
have
already considered has no effect on the equilibrium conditions (\ref{e14}).
Imagine some operator which is perfectly arbitrary except that it
conserves $\er$ number.  It can be characterized by a set of integers $n_F$,
the net number of fields of type $F$, defined to be the number of fields $F$
minus the number of charge-conjugated fields $F^c$.  If it is in equilibrium,
it gives a condition on the chemical potentials,
\beq
        \sum_F n_F \mu_F  = 0.
\label{e17}
\eeq
But the coefficients $n_F$ must be such that the total hypercharge of the
operator vanishes, because of gauge invariance:
\beq
        \sum_F Y_F n_F = 0.
\label{e18}
\eeq
Now note that if the solution for the chemical potentials takes the form $\mu_F
= c Y_F$ \cite{ahr}, then these equations are consistent with each other.  In
this case, eq.~(\ref{e17}) is automatically satisfied, so it is not really a
new condition, as we set out to demonstrate.  It is easy to verify from
eqs.~(\ref{e14}) that the chemical potential of each particle is indeed
proportional to its hypercharge. Therefore we can add as many new operators as
we want: as long as they conserve $\er$ number, they lead to the same
equilibrium conditions, which can be written as
\beq
        \mu_A = \frac{1}{20} Y_A \mu_\er.
\label{e19}
\eeq

\section{Summary and Conclusions}

Although the smallness of the electron Yukawa coupling is not completely
understood, its smallness is an experimental fact. In the absence of new
physics beyond the standard model and still below the GUT scale for which the
right-handed electron is not a singlet, $\er$ remains an essentially decoupled
state for a good part of the early evolution of the Universe. Indeed we have
accurately determined the decoupling temperature of this state to be $T_* =
1.3 f(x_\H)$ TeV.

Our interest in the $\er$ decoupling temperature is directly related to the
preservation and generation of the cosmological baryon asymmetry. We have shown
conclusively (up to mass effects \cite{krs2,dr}) that in a $B - L = 0$ universe
(with $B - L$ conserved) sphaleron erasure of a primordial asymmetry does
indeed occur in the standard model. The final asymmetry in this case is
exponentially sensitive to model parameters and it may be possible to preserve
the asymmetry in this case in an extension of the standard model.

Even in models which do not conserve $B-L$, sphaleron erasure of a baryon
asymmetry may occur if other $B/L$-violating interactions are included in any
extension of the standard model.  A simple example of such an interaction is
the $\Delta L = 2$ operator induced from neutrino mass generation via the
seesaw mechanism. The preservation of the baryon asymmetry places restrictions
on such operators and thus on the neutrino mass matrix.  In the literature
several widely differing bounds on neutrino masses have been given. Here  we
have clarified these bounds and shown again the importance of the $\er$
decoupling.  In particular we have shown that erasure may occur only between
$T_*$ and the electroweak phase transition temperature, $T_c$.  We have also
shown how the role of $\er$ decoupling should be applied to bounds on other
$B/L$ violating operators.

Finally, we have demonstrated that because of $\er$ decoupling, an initial
lepton or $\er$ asymmetry can be used to generate a baryon asymmetry even when
$B = 0$ initially. In such cases, not only do we drastically weaken upper
bounds on new operators, but even imply a {\it lower} bound such that they
would be able to induce a nonzero $B-L$ and ultimately the baryon asymmetry.

\vskip 1.0truecm
\noindent {\bf Acknowledgements}
\vskip 1.0truecm
We would like to thank Sonia Paban for useful discussions.
This work was supported in part by  DOE grant DE-AC02-83ER-40105.
The work of KAO was in addition supported by a Presidential Young
Investigator Award and the work of KK was in addition supported by
The Finnish Academy.
}}}

\newpage
\begin{center} {\large Figure Captions} \end{center}
\begin{description}

\item[Fig.~1] (a) Higgs boson decay
into right-handed electron and lepton doublet; Higgs and gauge boson scattering
of (b) $L_e$ into $e_R$ and (c) bosons into $L_e$ and $e_R$; lepton number
violating scatterings of (d) $e_L\ell^-$ or $e_L\nu_\ell$ into Higgs bosons and
(e) $e_L$-Higgs into $(\ell$ or $\nu_\ell)$-Higgs.

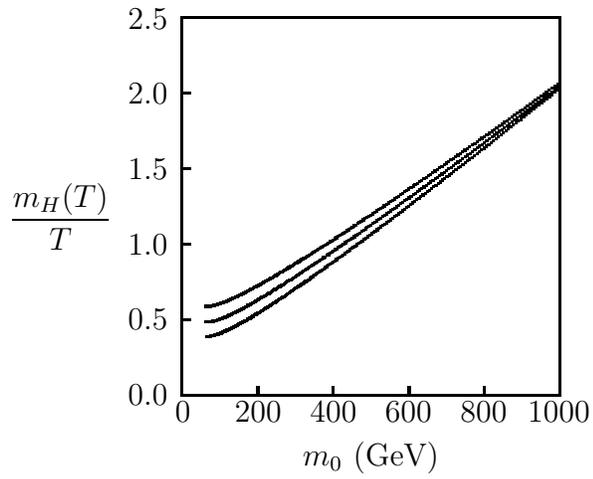
\item[Fig.~2] $m_H(T)/T$ as a
function of the zero temperature Higgs boson mass $m_0$ for top quark masses of
180 GeV (top curve), 135 GeV (middle curve) and 90 GeV (lower curve).

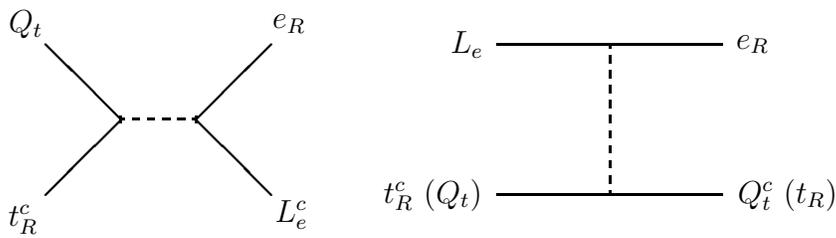
\item[Fig.~3] (a) $s$-channel
scattering of third-generation quark doublet and right-handed top antiquark
into $e_R$ and $L_e^c$; (b) $t$-channel scattering processes.

\item[Fig.~4] Diagrams contributing
to $e_R$ production by Higgs and gauge boson scattering.  The first two are
present for both $B$ and $W$ gauge bosons, whereas the last exists only for
$B$.

\item[Fig.~5] a) Contours of constant $B_{final}/B_{eq}$, when the scattering
contribution to the chirality changing rate is omitted. Solid lines correspond
to the present work; for comparison the previous result of CDEO is shown by the
dashed lines.  b) The solid lines are contours of constant $B_{final}/B_{eq}$
with the full interaction rate, including scatterings. Dashed lines are
contours of constant thermal Higgs mass parameter $x_H$.

\item[Fig.~6] a)  Evolution of the asymmetries $L_{e_R}$, $L_{e_L}$
and $B$ as a function of temperature, when electron lepton number is conserved
at low temperatures, using the initial value $L_{e_R,p} = 10^{-3}$.
b) Evolution of absolute values of asymmetries with electron lepton number
violation corresponding to $\langle m_{ei} \rangle = 5$ keV. $B$ is initially
negative and the dip in the corresponding curve marks the place where $B$
becomes positive.  Similarly $L_{e_L}$ changes sign from positive to negative.
$L_{e_R}$ remains negative over the whole range.
\newpage
\item[Fig.~7]  The shaded area between the curves C and D shows the region
consistent with the observed baryon asymmetry for some choice of parameters
$m_t$ and $m_H$. Curve C corresponds to $B,m_t,m_H=4\times10^{-11}$, 90 GeV,
60 GeV and curve D to $B,m_t,m_H=6\times10^{-11}$, 90 GeV, 1 TeV.

\end{description}
\newpage
\unitlength=1.00mm
\thicklines
\begin{picture}(143.66,142.34)
\put(80.33,128.67){\circle*{5.20}}
\put(70.33,118.67){\line(1,1){20.00}}
\put(70.33,138.67){\line(1,-1){20.00}}
\put(30.33,128.67){\circle*{5.20}}
\put(30.33,128.67){\line(1,1){10.00}}
\put(30.33,128.67){\line(1,-1){10.00}}
\put(130.33,128.67){\circle*{5.20}}
\put(120.33,118.67){\line(1,1){20.00}}
\put(120.33,138.67){\line(1,-1){20.00}}
\put(42.66,141.34){\makebox(0,0)[lc]{$e_R$}}
\put(42.66,115.34){\makebox(0,0)[lc]{$L_e^c$}}
\put(67.33,142.34){\makebox(0,0)[rc]{$L_e$}}
\put(93.33,141.67){\makebox(0,0)[lc]{$e_R$}}
\put(143.66,141.67){\makebox(0,0)[lc]{$e_R$}}
\put(143.66,116.34){\makebox(0,0)[lc]{$L_e^c$}}
\put(15.99,128.67){\dashbox{1.00}(11.67,0.00)[cc]{}}
\put(35.33,-32.00){\line(1,1){10.00}}
\put(45.33,-22.00){\line(0,1){0.00}}
\put(35.33,-32.00){\line(1,-1){10.00}}
\put(25.33,-32.00){\line(-1,1){10.00}}
\put(25.33,-32.00){\line(-1,-1){10.00}}
\put(75.33,-22.00){\line(1,0){30.00}}
\put(75.33,-42.00){\line(1,0){30.00}}
\put(90.33,-42.00){\dashbox{1.00}(0.00,20.00)[cc]{}}
\put(25.33,-32.00){\dashbox{1.00}(10.00,0.00)[cc]{}}
\put(13.00,128.67){\makebox(0,0)[cc]{$H$}}
\put(12.67,-19.33){\makebox(0,0)[cc]{$Q_t$}}
\put(12.33,-45.33){\makebox(0,0)[cc]{$t_R^c$}}
\put(47.67,-19.33){\makebox(0,0)[cc]{$e_R$}}
\put(48.00,-44.67){\makebox(0,0)[cc]{$L^c_e$}}
\put(71.33,-22.00){\makebox(0,0)[cc]{$L_e$}}
\put(73.33,-42.00){\makebox(0,0)[rc]{$t_R^c\ (Q_t)$}}
\put(109.33,-22.00){\makebox(0,0)[cc]{$e_R$}}
\put(107.33,-42.00){\makebox(0,0)[lc]{$Q^c_t\ (t_R)$}}
\put(20.00,0.33){\framebox(50.00,50.00)[cc]{}}
\bezier{236}(70.02,41.07)(27.76,7.46)(23.06,8.05)
\bezier{232}(70.02,41.36)(27.32,9.08)(22.91,9.96)
\bezier{228}(70.02,41.80)(28.05,11.57)(22.91,12.01)
\put(19.98,10.29){\line(1,0){1.03}}
\put(19.98,20.27){\line(1,0){1.03}}
\put(19.98,30.39){\line(1,0){1.03}}
\put(19.98,40.37){\line(1,0){1.03}}
\put(29.96,0.31){\line(0,1){1.03}}
\put(39.94,0.31){\line(0,1){1.03}}
\put(39.94,1.34){\line(-1,-3){0.15}}
\put(50.06,0.31){\line(0,1){1.03}}
\put(60.04,0.31){\line(0,1){1.03}}
\put(10.00,70.00){\makebox(0,0)[lt]{\parbox{6in}{\leftline{{\bf Figure 1}}
%(a) Higgs boson decay
%into right-handed electron and lepton doublet; Higgs and gauge boson
%scattering of (b) $L_e$ into $e_R$ and (c) bosons into $L_e$ and $e_R$;
%lepton number violating scatterings of (d) $e_L\ell^-$ or $e_L\nu_\ell$ into
%Higgs bosons and (e) $e_L$-Higgs into $(\ell$ or $\nu_\ell)$-Higgs.
}}}
\put(80.00,25.33){\makebox(0,0)[lc]{\parbox{3.25in}{{\bf Figure 2}  %$m_H(T)/T$
%%as a
%function of the zero temperature Higgs boson mass $m_0$ for top quark masses
%%of
%180 GeV (top curve), 135 GeV (middle curve) and 90 GeV (lower curve).
}}}
\put(130.00,-32.00){\makebox(0,0)[lt]{\parbox{1in}{{\bf Figure 3}  %(a)
%%$s$-channel
%scattering of third-generation quark doublet and right-handed top antiquark
%into $e_R$ and $L_e^c$; (b) $t$-channel scattering processes.
}}}
\put(18.00,10.33){\makebox(0,0)[rc]{$0.5$}}
\put(18.00,20.33){\makebox(0,0)[rc]{$1.0$}}
\put(18.00,30.33){\makebox(0,0)[rc]{$1.5$}}
\put(18.00,40.33){\makebox(0,0)[rc]{$2.0$}}
\put(18.00,0.33){\makebox(0,0)[rc]{$0.0$}}
\put(20.00,-2.00){\makebox(0,0)[cc]{$0$}}
\put(30.00,-2.00){\makebox(0,0)[cc]{$200$}}
\put(40.00,-2.00){\makebox(0,0)[cc]{$400$}}
\put(50.00,-2.00){\makebox(0,0)[cc]{$600$}}
\put(60.00,-2.00){\makebox(0,0)[cc]{$800$}}
\put(70.00,-2.00){\makebox(0,0)[cc]{$1000$}}
\put(18.00,50.33){\makebox(0,0)[rc]{$2.5$}}
\put(-3.00,24.00){\makebox(0,0)[lc]{$\displaystyle{m_H(T)\over T}$}}
\put(45.00,-6.00){\makebox(0,0)[ct]{$m_0$ (GeV)}}
\put(30.66,94.34){\circle*{5.20}}
\put(19.66,108.01){\makebox(0,0)[rc]{$e_L$}}
\put(19.66,81.01){\makebox(0,0)[rc]{$L_\ell$}}
\bezier{4}(30.66,94.34)(31.16,94.84)(31.67,95.35)
\bezier{4}(33.00,96.34)(33.50,96.84)(34.01,97.35)
\bezier{4}(35.01,98.35)(35.52,98.86)(36.02,99.36)
\bezier{4}(37.03,100.37)(37.47,100.81)(37.97,101.31)
\bezier{4}(38.97,102.31)(39.48,102.82)(39.98,103.32)
\bezier{4}(40.99,104.33)(41.49,104.83)(41.99,105.33)
\put(30.66,94.34){\line(-1,1){11.00}}
\put(30.66,94.34){\line(-1,-1){10.67}}
\put(80.66,94.34){\circle*{5.20}}
\put(69.66,108.01){\makebox(0,0)[rc]{$e_L$}}
\put(69.66,81.01){\makebox(0,0)[rc]{$H$}}
\bezier{4}(80.66,94.34)(81.16,94.84)(81.67,95.35)
\bezier{4}(83.00,96.34)(83.50,96.84)(84.01,97.35)
\bezier{4}(85.01,98.35)(85.52,98.86)(86.02,99.36)
\bezier{4}(87.03,100.37)(87.47,100.81)(87.97,101.31)
\bezier{4}(88.97,102.31)(89.48,102.82)(89.98,103.32)
\bezier{4}(90.99,104.33)(91.49,104.83)(91.99,105.33)
\put(80.66,94.34){\line(-1,1){11.00}}
\bezier{4}(69.66,83.00)(70.16,83.50)(70.67,84.01)
\bezier{4}(71.67,85.01)(72.18,85.52)(72.68,86.02)
\bezier{4}(73.69,87.03)(74.13,87.47)(74.63,87.97)
\bezier{4}(75.63,88.97)(76.14,89.48)(76.64,89.98)
\bezier{4}(77.65,90.99)(78.15,91.49)(78.65,91.99)
\put(91.66,83.34){\line(-1,1){11.00}}
\put(45.00,108.34){\makebox(0,0)[cc]{$H$}}
\put(44.66,80.00){\makebox(0,0)[cc]{$H$}}
\put(95.33,108.34){\makebox(0,0)[cc]{$H$}}
\put(95.33,81.34){\makebox(0,0)[cc]{$L_\ell$}}
\bezier{4}(41.30,82.99)(40.80,83.48)(40.31,83.98)
\bezier{4}(39.33,85.06)(38.84,85.55)(38.35,86.04)
\bezier{4}(37.36,87.02)(36.87,87.51)(36.38,88.01)
\bezier{4}(35.30,88.99)(34.81,89.48)(34.31,89.97)
\bezier{4}(33.33,91.05)(32.84,91.55)(32.35,92.04)
\end{picture}
\newpage

\begin{picture}(147.33,126.67)
\put(35.23,109.01){\line(1,1){10.00}}
\put(35.23,109.01){\line(1,-1){10.00}}
\put(25.23,109.01){\dashbox{1.00}(10.00,0.00)[cc]{}}
\bezier{20}(24.89,109.01)(25.04,106.45)(22.36,106.54)
\bezier{20}(22.36,106.54)(19.92,106.62)(19.92,104.02)
\bezier{20}(19.92,104.02)(20.01,101.42)(17.41,101.50)
\bezier{20}(17.41,101.42)(14.89,101.50)(14.89,98.99)
\put(80.33,103.68){\line(0,1){10.00}}
\put(80.33,113.68){\line(1,1){10.00}}
\put(80.33,103.68){\line(1,-1){10.00}}
\bezier{20}(80.33,103.65)(80.48,101.10)(77.80,101.18)
\bezier{20}(77.80,101.18)(75.36,101.26)(75.36,98.66)
\bezier{20}(75.36,98.66)(75.45,96.06)(72.85,96.15)
\bezier{20}(72.85,96.06)(70.33,96.15)(70.33,93.63)
\bezier{20}(130.00,103.66)(130.15,101.11)(127.46,101.19)
\bezier{20}(127.46,101.19)(125.03,101.27)(125.03,98.67)
\bezier{20}(125.03,98.67)(125.12,96.07)(122.52,96.16)
\bezier{20}(122.52,96.07)(120.00,96.16)(120.00,93.64)
\put(130.00,103.68){\line(2,3){14.33}}
\put(130.00,114.68){\line(2,-3){13.67}}
\bezier{4}(24.97,108.99)(24.47,109.48)(23.98,109.98)
\bezier{4}(23.00,111.06)(22.51,111.55)(22.02,112.04)
\bezier{4}(21.03,113.02)(20.54,113.51)(20.05,114.01)
\bezier{4}(18.97,114.99)(18.48,115.48)(17.98,115.97)
\bezier{4}(17.00,117.05)(16.51,117.55)(16.02,118.04)
\put(130.02,114.57){\line(0,-1){10.81}}
\bezier{4}(80.30,113.99)(79.80,114.48)(79.31,114.98)
\bezier{4}(78.33,116.06)(77.84,116.55)(77.35,117.04)
\bezier{4}(76.36,118.02)(75.87,118.51)(75.38,119.01)
\bezier{4}(74.30,119.99)(73.81,120.48)(73.31,120.97)
\bezier{4}(72.33,122.05)(71.84,122.55)(71.35,123.04)
\bezier{4}(129.97,114.66)(129.47,115.15)(128.98,115.65)
\bezier{4}(128.00,116.73)(127.51,117.22)(127.02,117.71)
\bezier{4}(126.03,118.69)(125.54,119.18)(125.05,119.68)
\bezier{4}(123.97,120.66)(123.48,121.15)(122.98,121.64)
\bezier{4}(122.00,122.72)(121.51,123.22)(121.02,123.71)
\put(12.33,120.67){\makebox(0,0)[cc]{$H$}}
\put(12.00,93.67){\makebox(0,0)[cc]{$B\ (W)$}}
\put(48.67,121.67){\makebox(0,0)[cc]{$e_R$}}
\put(49.33,94.34){\makebox(0,0)[rc]{$L^c_e$}}
\put(68.33,125.67){\makebox(0,0)[cc]{$H$}}
\put(68.00,89.01){\makebox(0,0)[cc]{$B\ (W)$}}
\put(93.00,126.01){\makebox(0,0)[cc]{$e_R$}}
\put(93.00,89.01){\makebox(0,0)[cc]{$L^c_e$}}
\put(118.33,126.34){\makebox(0,0)[cc]{$H$}}
\put(118.67,89.34){\makebox(0,0)[cc]{$B$}}
\put(147.33,126.67){\makebox(0,0)[cc]{$e_R$}}
\put(146.67,90.67){\makebox(0,0)[cc]{$L^c_e$}}
\put(60.67,109.01){\makebox(0,0)[cc]{$+$}}
\put(107.00,109.01){\makebox(0,0)[cc]{$+$}}
\put(10.00,81.34){\makebox(0,0)[lt]{\parbox{6in}{\leftline{{\bf Figure 4}}
%Diagrams contributing
%to $e_R$ production by Higgs and gauge boson scattering.  The first two are
%present for both $B$ and $W$ gauge bosons, whereas the last exists only for
%$B$.
}}}
\put(127.00,109.01){\makebox(0,0)[rc]{$e_R$}}
\put(83.33,109.01){\makebox(0,0)[lc]{$L_e$}}
\end{picture}
\vspace{ -2in}
\end{document}